\documentstyle[prl,epsf,aps]{revtex}
\newtheorem{theorem}{Theorem}[section]

\newtheorem{remark}{Remark} 
 
\newtheorem{lemma}[theorem]{Lemma}
\newtheorem{proposition}{Proposition}
\newtheorem{corollary}{Corollary} 
\newtheorem{conjecture}{Conjecture}
\newtheorem{assumption}{Assumption}

\newcommand{\bpr} {\noindent {\bf Proof }} 
\newcommand{\epr}{$\Box$}
\newcommand{\bd}{
\begin{document}} 
\newcommand{\ed}{\end{document}}
\newcommand{\beq}{\begin{equation}} 
\newcommand{\eeq}{\end{equation}}
\newcommand{\bef}{\begin{figure}} 
\newcommand{\enf}{\end{figure}}
\newcommand{\bea}{\begin{eqnarray}} 
\newcommand{\eea}{\end{eqnarray}}
\newcommand{\bth}{\begin{theorem}} 
\newcommand{\eth}{\end{theorem}}
\newcommand{\bhyp}{\begin{assumption}}
\newcommand{\ehyp}{\end{assumption}}
\newcommand{\bp}{\begin{proposition}}
\newcommand{\ep}{\end{proposition}}
\newcommand{\bco}{\begin{corollary}}
\newcommand{\eco}{\end{corollary}}
\newcommand{\bconj}{\begin{conjecture}}
\newcommand{\econj}{\end{conjecture}}
\newcommand{\ble}{\begin{lemma}} 
\newcommand{\ele}{\end{lemma}}
\newcommand{\bR}{\begin{remark}} 
\newcommand{\eR}{\end{remark}}
\newcommand{\bc}{\begin{center}} 
\newcommand{\ec}{\end{center}}
\newcommand{\ben}{\begin{enumerate}}
\newcommand{\een}{\end{enumerate}}
\newcommand{\bit}{\begin{itemize}} 
\newcommand{\eit}{\end{itemize}}
\newcommand{\su}{\section} 
\newcommand{\ssu}{\subsection}
\newcommand{\sssu}{\subsubsection} 
\newcommand{\nid}{\noindent}
\newcommand{\nnb}{\nonumber}

\newcommand\bbbr{{\sf I\!R}} 
\newcommand\bbbc{{\sf I\!C}}
\newcommand\bbbn{{\sf I\!N}} 
\newcommand\bbbh{{\sf I\!H}}
\newcommand\bbbz{{\sf Z\!\!Z}}

\newcommand\cA{{\cal A}} 
\newcommand\cB{{\cal B}} 
\newcommand\cC{{\cal C}} 
\newcommand\cD{{\cal D}} 
\newcommand\cE{{\cal E}}
\newcommand\cF{{\cal F}} 
\newcommand\cG{{\cal G}} 
\newcommand\cH{{\cal H}} 
\newcommand\cI{{\cal I}} 
\newcommand\cK{{\cal K}}
\newcommand\cL{{\cal L}} 
\newcommand\cM{{\cal M}} 
\newcommand\cN{{\cal N}} 
\newcommand\cP{{\cal P}} 
\newcommand\tcP{\tilde{\cal P}} 
\newcommand\cQ{{\cal Q}}
\newcommand\cR{{\cal R}} 
\newcommand\cS{{\cal S}} 
\newcommand\cW{{\cal W}} 
\newcommand\cU{{\cal U}} 
\newcommand\cV{{\cal V}}
\newcommand\cT{{\cal T}} 
\newcommand\cX{{\cal X}} 
\newcommand\cZ{{\cal Z}} 
\newcommand\cl{{\cal l}} 
\newcommand\cn{{\cal n}}

\newcommand\hX{\hat{\bX}} 
\newcommand\hY{\hat{\bY}} 
\newcommand\hF{\hat{\bF}}
\newcommand\hm{\hat{\mu}}
\newcommand\bE{{\bf E}} 
\newcommand\bF{{\bf F}} 
\newcommand\bJ{{\bf J}}
\newcommand\bX{{\bf X}} 
\newcommand\bY{{\bf Y}} 
\newcommand\bU{{\bf U}}
\newcommand\bZ{{\bf Z}} 
\newcommand\ba{{\bf a}} 
\newcommand\bb{{\bf b}} 
\newcommand\be{{\bf e}}
\newcommand\bk{{\bf k}}
\newcommand\bi{{\bf i}} 
\newcommand\bl{{\bf l}} 
\newcommand\bn{{\bf n}} 
\newcommand\bq{{\bf q}} 
\newcommand\bv{{\bf v}}
\newcommand\bx{{\bf x}} 
\newcommand\by{{\bf y}} 
\newcommand\bz{{\bf z}} 
\newcommand\bw{{\bf w}}
\newcommand\btX{{\bf \tilde{X}}} 

\newcommand\bcm{\bar{\cM}} \newcommand\cm{\cM}

\newcommand{\deq}{\stackrel {\rm def}{=}}

\newcommand{\sep}{\; \,} \newcommand{\D}{\displaystyle}
\newcommand{\T}{\textstyle} \newcommand{\etc}{etc $\dots$}
\newcommand{\etal}{etc $\dots$} \newcommand\dL{^\partial \Lambda}

\def\Appendix{\section*{APPENDIX}}

\baselineskip18pt

\bd
\draft

\baselineskip18pt

\title {Lyapunov exponents and transport in the Zhang model of
 Self-Organized Criticality.}

\author {B. Cessac
\thanks{Institut Non Lin\'eaire de Nice, 1361 Route des
Lucioles, 06560 Valbonne, France}
Ph. Blanchard
\thanks{University of Bielefeld, BiBoS, Postfach 100131, D-33501,
Bielefeld, Germany}
T. Kr\"uger
\thanks{University of Bielefeld, BiBoS, Postfach 100131, D-33501,
Bielefeld, Germany and Technische Universitaet, Str. des 17 July 135, 10623,
 Berlin, Germany}
}
\maketitle

\begin{abstract}
We discuss the role
played by the Lyapunov exponents in the 
dynamics of  Zhang's model of Self-Organized
Criticality. 
We show that
a large part of the spectrum (slowest modes) is associated with the energy transport
 in the lattice.  In particular, we give  bounds on the first negative Lyapunov
exponent in terms of the energy flux dissipated at the boundaries per unit of
time. We then establish an explicit formula for the transport modes
that appear as diffusion modes in a landscape where the metric is given
by the density of active sites. We use a  finite size scaling ansatz
for the Lyapunov spectrum
and  relate the scaling exponent to the scaling of quantities
like avalanche size, duration, density of active sites, etc ...
\end{abstract}

\pacs{PACS number: 02.10.Jf, 02.90+p, 05.45.+b, 05.40.+j} 
                      
%\pagebreak     

\su{Introduction.}

Within the last 10 years the notion of Self-Organized Criticality (SOC) has  become 
a new paradigm for the 
explanation of a huge variety of phenomena in nature and social sciences. Its 
origin 
lies in the 
attempt to explain the widespread appearance of power-law like statistics for
 characteristic events in a multitude 
of examples like the distribution of the size of earthquakes, 1/f-noise,
 amplitudes 
of solar flares,
species extinction .... to name only a  few cases \cite{BTW,Bak1,Jensen}. 
In this paradigm, 
the dynamics occur as chain reactions or avalanches.
Furthermore,  a stationary regime is reached, where
  the average incoming flux of external perturbations is 
balanced 
by the average outgoing flux that can leave the system at the boundary,
 or by dissipation in the bulk and
there is a constant flux through the system. 
In this stationary state,
referred to as the   \textit{SOC state},
the distribution of avalanches follows a power 
law, namely there is scale invariance reminiscent of thermodynamic systems at 
the 
critical point.  A local perturbation
can induce effects at any scale and there are 
long-range spatial and time correlations.
 In other words, in this paradigm 
the system  reaches {\it spontaneously} a
critical state without  any fine tuning of some control parameter.

Several models have been proposed to mimic this mechanism, like the sandpile
 model
\cite{BTW}, the abelian sandpile \cite{Dhar} or the continuous energy model 
\cite{Zhang}. The results available are mainly numerical and
only a few rigorous results are known.  
The numerical simulations report the following behaviour.
 Fix an  observable,
 say $x$, measuring
some property of an avalanche (duration, size, etc ...),
 and compute the related 
probability $P_L(x)$ at stationarity for a system of characteristic size $L$.
 The graph of $P_L(x)$ exhibits  a power law behaviour
over a finite range with a cut-off, corresponding to finite size effects. As $L$
increases the power law range increases, leading to the conjecture that
a critical state is indeed achieved in the thermodynamic limit,
namely that $P_L(x)$ behaves like $\frac{1}{x^{\tau_x}}$ as $L \to \infty$.
$\tau_x$ is called the {\it critical exponent} for the observable $x$.
There is apparently no control parameter to tune to attaining
 the critical state.
 Despite the large number of papers
written on the subject some basic problems are still open.

Guided by the wisdom coming from  renormalization group analysis 
and phase transitions in equilibrium systems,
 it seems natural to look for
a possible classification of the models into universality classes
characterized by a set of critical exponents, for a family of ``relevant''
avalanche-observables. However, the link 
between the ``criticality'' of the ``out of equilibrium''
SOC models  and the usual statistical mechanics of phase 
transitions in equilibrium systems remains to be clarified
\cite{Sornette}.
Furthermore, apart from the fact that the commonly studied
observables (size, duration, area, giration radius) do not necessarily
consitute a {\it complete} set allowing one to classify the models,
 even the computation of the critical 
exponents $\tau_x$ from numerical data 
is not easy and  there is no agreement yet on the way to do it.
It is clear that the simple measurement of the slope of $P_L(x)$ in the linear
range of a log-log plot is not reliable, due to the finite sample fluctuations
and because the explicit form of the cut-off is not known in general.
The computation of $\tau_x$ from the behaviour of the moments
is certainly a better way. However there is no agreement yet wether one should use 
a finite size scaling treatment \cite{Kadanoff} or  more sophisticated methods
 (like
multifractal analysis \cite{Tebaldi}). Therefore, at the moment, the identification
of a (supposed) universality class seems problematic. 
Finally, the central question is: what exactly 
does the knowledge of the 
critical exponents $\tau_x$ teach us about the model.?.

An alternative approach to better understand 
the behaviour of SOC models can also consist of studying the microscopic
dynamics  and to infer information about
 the macroscopic behaviour from this analysis. A detailed analysis
 can, at first
sight, seem 
useless since the conventional wisdom from classical
statistical mechanics is that microscopic ``details'' are irrelevant,
and only structural properties like conservation laws and symmetries
are essential. However, as mentionned above, the theory of SOC
has not yet reached the level of understanding
of classical critical phenomena. It suffers in particular
from the lack of a thermodynamic formalism and notions like Gibbs measures
and free energy can a priori not be used. On the other hand,
by having a precise description of the dynamics of the finite size system
one can expect a better understanding of the thermodynamic limit
and decide which components in the models definition are really ``relevant'',
and which information does the usually computed quantities (like
critical exponents) actually give us.

This is the essence of the program we developped in \cite{BCK1,BCK2,BCK3}.
We found that  Zhang's model of SOC \cite{Zhang}
can be fruitfully studied with the  tools of
 hyperbolic dynamical system theory.
Then we were able to extract unexpected results,
establishing in particular a formula relating the critical exponent
of avalanche size to the spectrum of the Lyapunov exponents.
 In this paper we develop this point of view and  make a further step
 towards unterstanding  the dynamical properties of this model
and their link to the SOC state.
We first define the model as a hyperbolic dynamical system of skew-product
type.  We then define in this setting
 two different time scales : the {\it local time} which is the natural time
for the dynamical system, and the {\it avalanche time}, related to the
avalanche duration.
We introduce a natural invariant measure to characterize
the statistical properties at stationarity, and we relate the avalanche
 observable
statistics to the ergodic local time average. We then discuss the role
played by the Lyapunov exponents in the dynamics and their relation to the 
energy transport and to the average avalanche observables. We show that random
 excitation
induces a positive Lyapunov exponent, while the relaxation dynamics
corresponds to negative exponents. Furthermore, we show that
a wide part of the spectrum (slowest modes) is associated with  energy transport
 in the lattice. In particular, we give  bounds on the first negative Lyapunov
exponents in terms of the energy flux dissipated at the boundaries per unit
of time. We  establish an explicit formula for the transport modes,
that appear as diffusion modes in a landscape where the metric is given
by the density of active sites. Except for the first modes, they differ dramatically
from the normal diffusion modes one would obtain by assuming a uniform density of
active sites. It has been argued in  \cite{Vespignani}
that SOC requires a wide separation between the excitation and relaxation
time scales (slow driving). We actually show in this paper,
as a consequence of our more general analysis,
that the dynamics of the Zhang provides naturally this separation,
and that  the \textit{infinitely slow driving} limit
is actually reached as the size of the system goes to infinity.
 We then show that, using  a finite size scaling ansatz for
 the Lyapunov spectrum, one can relate the obtained scaling exponent 
to the scaling of quantities
like avalanche size, duration, density of active sites, etc ....

\su{Dynamical system definition  and basic properties.}

\ssu{Definition.}

The Zhang's model \cite{Zhang}, widely inspired from the Bak-Tang-Wiesenfield
precursor model \cite{BTW}, has been introduced as a possible example of a
model which ``self-organizes'' into a critical state in the thermodynamic
limit, namely without  fine-tuning of a control parameter.
Its beauty lies in its simplicity.

Let $\Lambda$ be a d-dimensional sub-lattice in $\bbbz^d$, taken as a
square of edge length $L$ for simplicity. Call $N=\#\Lambda=L^d$, and
let $\dL$ be the boundary of $\Lambda$, namely the set of points in
$\bbbz^d \setminus\Lambda$ at distance $1$ from $\Lambda$.  Each site
$i \in \Lambda$ is characterized by its "energy" $X_i$, which is a
non-negative real number. Call  $\bX = \lbrace X_i
\rbrace_{_{i \in \Lambda}}$ a configuration of energies.
Let $E_c$ be a real, strictly positive number, called the {\it critical energy},
and
$\cm = [0,E_c[^N$. A configuration
$\bX$ is called "stable" iff $\bX \in \cm$
and "unstable" otherwise. If $\bX$ is stable then
one chooses a site $i$ at random with some probability $\nu_L(i)$, and add to
 it energy
$\delta$, where $\delta$ is set to 1 in this paper (excitation).
If a site $i$ is \textit{overcritical} or \textit{active}
 ($X_i \geq E_c$), it loses a part of its energy
 in equal parts to its $2d$ neighbours (relaxation).
 Namely, we fix a parameter
$\epsilon \in [0,1[$ such that, after relaxation of the site $i$,
the remaining energy of $i$ is $\epsilon X_i$, while the $2d$ neighbours
receive the energy $\frac{(1-\epsilon)X_i}{2d}$. Note therefore that there  is
{\it local conservation of energy}. 
If several nodes are simultaneously active,
the local distribution rules are
additively superposed,
i.e. the time evolution of the system is synchronous.  The succession
of updating leading an unstable configuration
to a stable one is called an {\it avalanche} (a more precise definition
of an avalanche will be given below).   
 There  is dissipation at the boundaries namely
the sites of $\dL$ have always
zero energy. As a result, all avalanches are {\it finite}.
The addition of energy is {\it adiabatic}. When an avalanche occurs,
one waits until it stops before adding a new energy quantum.
Further excitations eventually generate a new avalanche,
but, because of the adiabatic rule, each new avalanche starts from {\it only one}
active site. Note that relaxation depends on {\it local} 
conditions while excitation
is conditioned by {\it global} constraints (all sites are quiescent).
It is conjectured that a critical state is reached, independently of $E_c$,
at least for large $E_c$ values \footnote{Strong deviations from a power law
have been observed for small $E_c$ in one dimension \cite{BCK1}.}.

\ssu{The Zhang's model as a dynamical system.}

Because 
all avalanches are {\it finite} (for finite $L$), and
since we are not interested in the  transients, one can, without loss of
generality take all initial energy configurations $\bX \in \cm$. All
 trajectories 
starting from $\cM$ belong to a compact set $\cB$.
 Call $\bar{\cM}= \cB \setminus \cM$. $\bar{\cM}$ contains  the set of all 
unstable
energy configurations achievable in an avalanche starting from an
energy configuration in $\cM$.

Fix $\epsilon >0$, and call $\alpha=\frac{1-\epsilon}{2d}$.  Let
$h$ be the Heaviside function.  Define $H: \bbbr^N \rightarrow
\left\{0,1\right\}^N$ such that $H(\bX)$ is the vector $\left\{h(X_i)
\right\}_{i=1 \dots N}$. Call $\bX_c$ the vector $\left\{E_c \right\}_{i=1
 \dots N}
$. Finally, let $\Delta$ be the discrete Laplacian. The dynamics is defined
by the mapping  $\bF : \cB \to \cB$ such that:
\beq\label{relaxation}
 \bF(\bX)=\bX + \alpha\Delta\left[H(\bX-\bX_c).\bX \right] 
\eeq
\nid which redistributes the energy of the active sites in equal parts to the
 neighbours
after one relaxation
step. Note that $\bF$  is 
the identity if no site is active, i.e. if $\bX \in \cm$,
and that it is \textit{piecewise linear},
(i.e. linear on sub-domains $\cB_k \in \cB$).
$\bF$ is a (singular) diffusion operator and $\alpha$  the diffusion
coefficient.
 
It is useful to encode the dynamics of excitation in the following way.
 Let
$\Sigma^+_\Lambda$ be the set of right infinite sequences $\ba =
\left\{ a_1,\dots, a_k, \dots\right\}, a_k \in \Lambda$, and $\sigma$
be the {\it left shift} over $\Sigma^+_\Lambda$, namely
$\sigma \ba = a_2a_3 \dots $. The elements of $\Sigma^+_\Lambda$ are called
{\it excitation sequences}.
The set $\Omega=\Sigma^+_\Lambda \times
\cB$ is the {\it phase space} of the Zhang's model 
and $\hX= (\ba,\bX)$ is a point in
$\Omega$.  The Zhang's model dynamics is given by a map of
skew-product type $\hF : \Omega \rightarrow \Omega$ such that:
\bea \bX \in \cm \Rightarrow \hF(\hX) &=&
\left(\sigma.\ba,\bX+\be_a \right)\\
\bX \in \bcm \Rightarrow \hF(\hX) &=& \left(\ba,\bF(\bX)\right) \eea

The knowledge of an initial energy configuration $\bX$, and of an
(infinite) sequence of excited sites $\ba$ (resp.
of an initial condition $\hX$) fully determines the evolution.
One can give $\Sigma^+_\Lambda$ a probability distribution $\nu_L$
corresponding
to  a random choice for the excited sites.
In the original Zhang's model the excited sites are choosen at random and 
independently with
 uniform probability. This corresponds to giving $\Sigma^+_\Lambda$ 
the \textit{uniform
Bernoulli measure}. Throughout this paper we will often 
 think of the left
Bernoulli shift on $\Sigma^+_\Lambda$ as represented by the
 system $z \to N.z \ \mbox{mod}\ 1, z \in [0,1]$.

In the following we will denote the two projections on the first and 
second coordinate by
$\pi^u(\hX) = \ba$, and $\pi^s(\hX) = \bX$.
 The superscripts $u,s$ mean respectively {\it
  unstable} and {\it stable} and correspond to the expansion (resp.
contraction) properties of the dynamics. Let
$D\hF_{\hX}$ be the tangent map of $\hF$ at $\hat{\bX}$
and $D\hF^t_{\hX}$ the $t$-th iterate. As shown below
$\pi^u(D\hF_{\hat{\bX}})$ is expansive whereas
$\pi^s(D\hF_{\hat{\bX}})$ induces  contraction.
In the following we will use the notation 
$\hX(t) = \hF^t(\hX)$
(resp. $\bX(t) = \pi^s(\hF^t(\hX))$). Furthermore note that
$\pi^s(D\hF_{\hat{\bX}})=D\bF_{\bX}$, and that $D\bF_{\bX}=I$, the identity
matrix over $\bbbr^N$, if $\bX \in \cm$.  \\
    
Consider a point $\hX \in \Omega$. Its trajectory is intermittent,
composed of bursts of excitation of the sites $a_1, a_2, \dots a_n$,
for those times $t$ such that $\bX(t) \in \cm$, followed by relaxation
periods when $\bX(t) \in \bcm$. 
 Define  the following hierarchy of \textit{waiting times}:
\bea
\gamma_0(\hX)&=&0\\
\sigma_i(\hX)&=&\inf_{t>\gamma_{i-1}} \left\{ \bX(t) \in \bcm  \right\}, 
\quad i \geq 1\\
\gamma_i(\hX)&=&\inf_{t>\sigma_{i}} \left\{\bX(t) \in \cm  \right\}, \quad 
i \geq 1 
 \eea
For $i\geq 1$, $\sigma_i(\hX)$ (resp. $\gamma_i(\hX)$ ) is the
starting time (resp. ending time) of the $i$-th avalanche occuring
during the evolution of $\hX$. Therefore the {\it avalanche
duration} of the $i$-th avalanche is :
\beq \tau_i(\hX)= \gamma_i(\hX) - \sigma_i(\hX) \eeq

In the same way, one defines :
\beq \omega_i(\hX)= \sigma_i(\hX) - \gamma_{i-1}(\hX) \eeq
\nid which is the number of excitations between the end of the
avalanche $i-1$ and the beginning of the next avalanche. 
In this way, one introduces naturaly two  time scales :
the \textit{local time} $t$ corresponding to one step of iteration in
the dynamics, and the \textit{avalanche time} $\tau_i$ corresponding to the
duration of an avalanche (a similar description has been used in \cite{Diaz}).

The waiting times are  useful for defining the usual avalanche observables. 
The number of relaxing sites for a given configuration is :
\beq 
r(\hX) = \#\left\{i \in \Lambda, \ X_i \geq E_c \right\} 
\eeq
The \textit{avalanche size} is
\beq 
s(\hX) = \sum_{t=1}^{\tau(\hX)} r(\hF^t(\hX)) 
\eeq
\nid where:
\beq 
\tau(\hX) = \inf_{t \geq 1} \left\{\bF^t(\bX)\in \cm \right\}-1 
\eeq
\nid is  duration of the avalanche occuring when exciting the site
$a_1$ in a stable energy configuration $\bX$. It is zero if one drops energy
without relaxation.

The structure of an avalanche  can be encoded by the sequence of
active sites $A(\hX) = \left\{A_t(\hX)\right\}_{1 \leq t \leq \tau(\hX)}$ 
where $A_t(\hX) = \left\{j \in \Lambda | X_j(t) \geq E_c \right\}$.
(Note that $A_1(\hX)$ is non empty and equals $\left\{a_1\right\}$ iff 
$\bX+\be_{a_1}$ is active). 
Correspondingly, there exists a partition\footnote{This partition is induced
by the partition of $\cB$ into domains of continuity for $\bF$ \cite{BCK3}.}
 of $\Sigma^+_\Lambda \times \cm$ 
into domains
$\cP_{i,k} = [i]\times \cm_{i,k}$ 
where $[i]$ is the set of sequences in $\Sigma^+_\Lambda$
having a first digit $i$,
such
that for any energy configuration $\bX \in \cm_{i,k}$
  the excitation of the site $i$
leads to the same avalanche (namely the same sites relax at the same time).
Under some moderate assumptions (see  \cite{BCK3}),
this allows us to define a symbolic coding for the avalanche and a transition
graph giving the transition rules between successives avalanches,
and to show that the
 dynamical
system admits a unique, fractal, invariant set. The boundary of the domains
 $\cP_{i,k}$
constitutes the \textit{singularity set} of $\hF$, called $\cS$.
This is the set of points where
$\hF$ is not continuous.
  
\ssu{Stationary state and probability of avalanche observables.}

Let $\hm_L$ be an invariant measure for the dynamical
system $\left\{\Omega,\hF \right\}$, where $L$ refers to the lattice size,
 namely 
$\hm_L(\hF^{-1}(\cA))=\hm_L(\cA)$
where $\cA \in \Omega$ is a measurable set.  
Since $\Omega$ has a product structure where the unstable
foliation is always transverse to the stable one,
 and since the dynamical system is
a skew product,
$\hm_L=\nu_L \times \mu_L $,
where  $\nu_L$ is the 
induced measure on
the unstable direction or \textit{excitation} measure,
 and $\mu_L$ is the induced measure on $\cB$ or measure on the \textit{energy}
 configurations. 
For simplicity we will assume
that $\nu_L$ is a Bernoulli measure, namely that the successive excited
sites are chosen \textit{independently}
with fixed rates.
Once we have fixed the distribution
 of excitation, we  are interested on the possible $\mu_L$ measures.
Of special physical importance are the measures obtained by iterating the 
Lebesgue measure $\mu_{Leb}$\footnote{Or any absolutely continuous measure,
which corresponds to selecting the initial energy configuration
with a probability distribution having a density.} 
on $\cm$, that is 
$\lim_{n \rightarrow \infty} 
\frac{1}{n}\sum_{i=0}^{n-1} \hF^{i}(\nu_L \times \mu_{Leb})$.
When the excitation measure 
$\nu_L$ is itself the Lebesgue measure on $[0,1]$ (corresponding
to choosing the excited sites with  uniform
probability) the measure obtained is called the Sinai-Ruelle-Bowen
measure (SRB). More generally, we will call (conditional) SRB the measure
$\lim_{n \rightarrow \infty} 
\frac{1}{n}\sum_{i=0}^{n-1} \hF^{i}(\nu_L \times \mu_{Leb})$, for a fixed
 $\nu_L$.
This is a natural  invariant measure from the physicists point of 
view since it gives the ensemble average with respect to typical  
initial energy configurations.
 
It is common in the SOC literature to assume ergodicity. 
In our setting, the physically relevant 
ergodic property is equivalent to assuming that the SRB measure is unique.
Proving the ergodicity in the Zhang's model,
 is clearly a difficult task which
is beyond the scope of this paper. We note however that this point
has been widely discussed in a previous paper \cite{BCK3},
where strong mathematical arguments in favour of this were given.
Actually, ergodicity was proved, but 
restricted to the one dimensional model,
and to some $E_c$ interval. A general proof is under construction
and will be published elsewhere \cite{BCK4}.
On physical grounds, note \textit{a contrario} that the failure
of ergodicity would lead to a stationnary state depending on initial
conditions. This would contradict the implicit SOC assumption that
the stationnary state is unique. In the following, we will therefore assume
that ergodicity holds and that  $\hm_L$ is the unique SRB measure.
This implies in particular the almost-sure equality between the ensemble average
 and 
the time average, namely, if $\phi$ is some observable, (a function $\Omega 
\to \bbbr$, integrable with respect to $\hm_L$) 

\beq
 \bar{\phi}_
L \deq
\lim_{T \rightarrow \infty}
\frac{1}{T}\sum_{t=1}^{T} \phi(\hF^{t}(\hX))=\int_\Omega\phi(\hX)d\hm_L(\hX)
\deq E_L[\phi]
\eeq

\nid for a typical (namely
Lebesgue almost surely) initial condition  $\hX$. Here, and in the following
$\bar{}_L$ will denote the time average, while $E_L[]$ will be
the ensemble average, on a lattice of size $L$.\\

From the dynamical systems point of view $\hm_L$ is the natural object
to deal with. However, in the SOC literature one is more interested in the 
probability distribution
 of some avalanche observable and its 
 scaling properties  in the thermodynamic limit. 
Fix an avalanche observable, say $s$. Call $\cP_s$ the union of
domains $\cP_{i,k}$ such that the avalanche corresponding to
each domain $\cP_{i,k}$ 
has the same size $s$.  Then the probability of having
an avalanche of size $s$, by excitation of a {\it stable} configuration, is 
$
Prob[s(\hX)=s | \hX \in \cP] =
 \frac{\hm_L(\cP_s)}{\hm_L(\cP)} =\frac{\hm_L(\cP_s)}{\mu_L(\cm)}
$.
In this definition we include the avalanches of size zero (excitation
without relaxation).
However, it is more natural from the SOC point of view to exclude this
case. We therefore define 
$P_L(s)$ as the probability  of having an avalanche of size $s$ 
{\it strictly larger than 0}\footnote{In view of the expected
critical behaviour as $L\to \infty$, one usually 
writes a scaling form $P_L(s) =\frac{f_L(s)}{s^{\tau_s}}$ 
where $f_L(s)$ is a cut-off term accounting for finite
size effects on large scales. $P_L(s)$   is not defined for
$s=0$ unless assuming very special properties for $f_L(s)$.}. 

\beq
P_L[s] \deq \frac{\hm_L(\cP_s)}{p_L}, \quad s \geq 1
\eeq

\nid where $p_L  \deq Prob\left[s(\hX) \geq 1 , \bX \in \cm \right]$ 
is the probability of \textit{initiating} an avalanche. 
The average with respect to $P_L[s]$, denoted further on by $< >_L$, is  :
\beq
<\psi(s)>_L \deq \sum_{s=1}^{\xi_L^s} P_L[s]\psi(s)  
\eeq

\nid where $\psi$ is some real function,
 and $\xi_L^s$ is the maximal value
that the observable $s$ can have on a lattice of size $L$ (note that $\xi_L^s$
depends also on $E_c,\epsilon,d$ but is \textit{bounded} if $L<\infty$).
The same definition holds for any other avalanche observable.
From the ergodic theorem:

\beq
<\psi(s)>_L=
\lim_{n \to \infty} \frac{1}{n} \sum_{i=1}^{n} \psi(s_i)
\eeq

\nid where $s_i$ is the size 
of the $i$-th avalanche occuring in the trajectory of
a generic point $\hX$.

One has: 

\beq \label{pL}
p_L = \D{\hm_L\left[\bigcup_{i=1}^N \left\{a_1=i,X_i \in [E_c-1,E_c[ \right\}
 \right]} =
 \sum_{i=1}^N  p_L(i) 
\eeq

\nid where:

\beq
p_L(i) \deq \nu_L(i).\mu_L\left\{X_i \in [E_c-1,E_c[\right\}
\eeq

\nid is the probability that \textit{an avalanche starts  at the site $i$}.
 Note that the probabilities $p_L(i)$
{\it depend} a priori on $i$ even if the excitation measure is uniform.
In this case, however, (\ref{pL}) reduces to

\beq
p_L = \frac{1}{N}.\sum_{i=1}^N \mu_L\left\{X_i \in [E_c-1,E_c[\right\}
\eeq

 Fix $\hX$ and $T$, then
call $n(T,\hX)$ the number of \textit{complete} avalanches occuring until 
local time
$T$ for the initial condition $\hX$. Obviously,
 $n(T,\hX) \to \infty$ as $T \to \infty, \ \forall \hX$. Then
from the ergodic theorem :
\beq
p_L = \lim_{T \rightarrow \infty} \frac{n(T,\hX)}{T} 
\eeq 

One can decompose $T$ has : $\D{T=\sum_{i=1}^{n(T,\hX)}\tau_i
+ \sum_{i=1}^{n(T,\hX)}  \omega_i
+K(\hX)}$ where $K(\hX)$ is some  residual time, finite, whatever $T$,
whatever $\hX$ ($K(\hX)$ is bounded by the largest avalanche duration).
Note that $\tau_i$ (resp. $\omega_i$) stands for $\tau_i(\hX)$ 
(resp. $\omega_i(\hX)$)
 but we removed the $\hX$ dependence
in order to simplify the notations.
 Then, as $T$ goes to infinity:

$$\D{\frac{n(T,\hX)}{T} \sim \frac{n(T,\hX)}{\sum_{i=1}^{n(T,\hX)}\tau_i +
 \sum_{i=1}^{n(T,\hX)} 
 \omega_i}} 
\D{ = \frac{n(T,\hX)}{\sum_{i=1}^{n(T,\hX)}\tau_i}-\frac{n(T,\hX)}
{\sum_{i=1}^{n(T,\hX)}\tau_i}
\frac{\sum_{i=1}^{n(T,\hX)}  \omega_i}{T}}$$

Call:

\beq \label{bom}
\bar{\omega}_L \deq \lim_{T \rightarrow \infty} 
\frac{1}{T}.\sum_{i=1}^{n(T,\hX)}  \omega_i
= \mu_L(\cm)
\eeq
 
\nid the {\it probability of dropping  energy in the system, at a given time},
(the equality holds for $\mu_L$ almost-every $\hX$ from the ergodic theorem).
$\bar{\omega}_L(i)=Prob[a_1=i,\bX \in \cm]=\nu_L(i)\bar{\omega}_L$,
is the probability of dropping energy on the site $i$, at a given time,
and  is  called
the {\it driving rate} in the literature \cite{Vespignani2}.   
One has:

\beq \label{plomega}
p_L = \frac{1-\bar{\omega}_L}{<\tau>_L} = \frac{\mu_L(\bcm)}{<\tau>_L}
\eeq

\nid where $<\tau>_L$ is the \textit{average avalanche duration}.

\bigskip

There exists an important relation linking the avalanche averages
(average with respect to $P_L$) to the local time average (average 
with respect to
$\hm_L$). Let $\phi : \Omega \to \bbbr$ be some observable
{\it such that $\phi(\hX) = 0$ whenever $\bX \in \cm$}.
A related avalanche observable can be defined by summing the
values that $\phi$ takes in one avalanche. Namely,
call $f_i(\hX) = \sum_{t=\sigma_i(\hX)}^{\gamma_i(\hX)} \phi(\hX(t))$.
(An important example is when $\phi(\hX)=r(\hX)$, the number
of active sites in one step. Then $f_i(\hX)$ is the size
of the $i$ th avalanche in the trajectory of $\hX$). One obtains:

$$
\bar{\phi}_
L =
\lim_{T \rightarrow \infty}
\frac{1}{T}\sum_{i=1}^{n(T,\hX)}\sum_{t=\sigma_i(\hX)}^{\gamma_i(\hX)} 
\phi(\hX(t)) 
$$

\nid which yields:

\beq \label{tvslent}
\bar{\phi}_L = p_L.\left<f\right>_L
\eeq

In particular :

\beq
\bar{r}_L = p_L.\left<s\right>_L
\eeq

Finaly we define the probability that a site $i$ is active
(often called the {\it density of active sites} in the literature 
\footnote{We will keep
this terminology throughout this paper though $\rho_L(i)$ is not strictly 
speaking 
a density
since $\sum_{i=1}^N \rho_L(i) \neq 1$.}): 

\beq
\rho_L(i) \deq \mu_L\left[X_i \geq E_c \right]
\eeq

\nid and

\beq
\rho_L^{av}=\frac{1}{N}\sum_{i=1}^{N}\rho_L(i) 
\eeq
 
$\rho_L^{av}$ is believed to act as an order parameter in the Zhang's model.

\su{Dynamical properties and Lyapunov exponents.}

\ssu{Jacobian matrix and Lyapunov exponents.}

Due to the piecewise affine structure of the map
$\bF$, the Jacobian matrix $DF_{\bX}$ plays a central
role in the Zhang's model, since it characterizes the energy
transport. Indeed,
the entry  $DF^t_{\bX,ij}$  is the ratio of energy 
flowing  from site $j$ to site $i$ in $t$ times steps for the initial condition
 $\bX$. Define 
$Z_k(\bX(t))=H(X_k(t)-E_c)$. This is a random variable,
taking value 0 if $X_k(t)$ is stable, and value 1 otherwise, whose
probability distribution is induced (at stationarity) by the invariant
measure $\hm_L$. More precisely, $Prob[Z_k(\bX(t))=1]=\rho_L(k)$.
 Let $\bZ(\bX)=\left\{Z_k(\bX) \right\}_{k=1}^N$,
and call $S(\bX)=\Delta\bZ(\bX)I$ (equivalently
$S(\bX)$ is the matrix of entries $S_{ij}(\bX)=\Delta_{ij}Z_j(\bX)$).
$S$ is the ``toppling'' operator of the Zhang's model.
 The jacobian matrix writes $DF_{\bX}=I+\alpha.S(\bX)$, while
  $DF^t_{\bX}$ is given by :

\bea \label{DFt}
DF^t_{\bX} = I + \alpha\sum_{t_0=1}^tS(\bX(t_0))
+\alpha^2\sum_{t\geq t_1 > t_0\geq 1}S(\bX(t_1))S(\bX(t_0)) &+& \dots \\\nonumber
+\alpha^r\sum_{t\geq t_{r-1}> t_{r-2} \dots  > t_0\geq 1}S(\bX(t_{r-1}))
S(\bX(t_{r-2})) 
\dots S(\bX(t_0))
&+& \dots
+\alpha^tS(\bX(t))S(\bX(t-1))\dots S(\bX(1))
\eea

Therefore, the generic term (say of order $r$) is a a ``propagator'' 
transmitting
the energy in r times steps. Note that this formula is exact.
It calls for the following remarks:

\bit
\item The maps $S(\bX)$ do not commute, and they depend on the state.
This is a key difference from the Dhar's model since it induces a 
\textit{non abelian}
structure and a ``toppling'' operator depending \textit{not only on the site,
but also on the whole energy configuration}. In particular the propagator
\textit{is not a mere polynomial of the Laplacian}.

\item The evolution depends \textit{a priori} on the whole trajectory and 
therefore
the strong memory effects expected in a critical phenomenon, can 
be treated from eq. (\ref{DFt}).
\eit

If one defines the excitation times for a given trajectory by:

\beq
\nu_k(\hX)=\inf_{t>\nu_{k-1}(\hX)}\left\{\bX(t) \in \cm\right\}
\eeq

\nid with $\nu_0=\gamma_0=1$, the energy configuration 
at time $T$, for an initial condition $\hX$ is:

\beq \label{X(T)}
\bX(T) =  DF_{\bX}^T.\bX + \sum_{i=1}^{m(T,\hX)} 
DF_{\bX}^{t-\nu_i(\hX)}.\be_{a_{\nu_i(\hX)}}
\eeq 

\nid where $m(T,\hX)$ is the number of excitations on a time interval
of length  $T$ 
for the initial condition
$\hX$. The first term corresponds to the 
redistribution of the initial energy configuration while
the second one
corresponds to the
redistribution of the energy quantum $\delta=1$ dropped in the system at times 
$\nu_k(\hX)$. Since the equilibrium average is assumed to be independent
of the initial condition, the first term has to decay
to zero as  $t \to \infty$, with a decay rate corresponding to the characteristic
relaxation time to equilibrium.
 
It is  therefore  important to well understand
the (spectral) properties of the $DF^t{\bX}$ in the infinite time limit. 
Were $S(\bX)$ be the Laplace operator, were the spectrum of 
$DF^t{\bX}$ be composed by Fourier modes,
and the relaxation time to equilibrium would be the slowest mode.
However, the mere presence of a singular term $Z(\bX)$ certainly makes
a big difference. Since $S$ depends on $\bX$ one clearly has to
study the decay rates  averaged on a full (typical) trajectory
or equivalently to compute the ensemble average.  In this view,
the law of the stochastic process $\left\{Z(\bX(t))\right\}_{t=0}^{+\infty}$
 (namely the density of active sites and correlations
at all 
times) certainly plays a role.

 The numbers
characterizing  the decay (resp. expansion) rates of the norm of
a small
pertubation in the trajectory's tangent space  of a
point $\hX$ under the action of the infinite
product matrix $DF^t{\hX}, t \to \infty$ are the \textit{Lyapunov
exponents}. They are \textit{real} numbers, well defined under some moderate
assumptions on $DF{\hX}$ (see \cite{Oseledec}) and are almost-surely
independent of $\hX$. Furthermore they are also independent of the norm
(in finite dimension). 

As shown in \cite{BCK3} and widely discussed in this paper
all the Lyapunov exponents are  different
from zero, for \textit{finite} $L$ (weak hyperbolicity).
One remarkable consequence is that the asymptotic dynamics
lies onto a fractal attractor and that the Lyapunov spectrum is closely related
to the (local) fractal properties of the invariant set through
the Kaplan-Yorke \cite{ER} and the Ledrappier-Young formula \cite{LY,BCK3}.
At this point a remark is  necessary.
Hyperbolicity  is clearly  a particular feature 
of  Zhang's model and of similar models where
 the amount of redistributed
energy from a critical site $i$ depends on its energy $X_i$.
By opposition, in models like
 BTW's or  Dhar's model the amount of transfered energy is a constant.
As a direct consequence, in these models, the dynamics is simply a 
piecewise translation
in the phase space, and  the uniform measure
is preserved \cite{Dhar} 
Hence,
 all lyapunov exponents are zero \cite{Erzan}. There is therefore  clearly
a \textit{structural difference} between the dynamics of Zhang's type models
and sandpiles. This observation seems at first sight to ruin the hope
to classify the Zhang's type models and the sanpiles in the same
``universality class''. However, we show in this paper that
hyperbolicity of  Zhang's model is partially lost in the thermodynamic
limit. Namely, some of the Lyapunov exponents go to zero as $L\to \infty$,
with a polynomial rate (exponent $\tau_\lambda$ in the last section)
closely related to SOC critical exponents. It might therefore
well be that these two class of model share the same SOC
critical exponents in the thermodynamic limit, though their
dynamics are still of different nature, even in this limit.\\

Due to the skew product structure,  the tangent map at any point
$\hX$ admits
a natural splitting $D\hF=(\pi^u(D\hF_{\hX}),\pi^s(D\hF_{\hX}))$
where the one dimensional map
$\pi^u(D\hF_{\hX})$ is expansive. Indeed,  the average expansion rate is given
by :

\beq
\lambda_L(0)=\lim_{T \rightarrow \infty} \frac{1}{T} 
log(det(\pi^u(D\hF^T_{\hX})))
=\bar{\omega}_L.log(N)
\eeq  

\nid since $\D{det(\pi^u(D\hF^T_{\hX}))=N^{\sum_{i=1}^{n(T,\hX)}\omega_i(\hX)}}$.
Therefore, since $\bar{\omega}_L \neq 0$, there is a {\it positive Lyapunov 
exponent}
in the dynamics. Note that this is  due to the excitation rule, and that it
reflects simply the ``chaotic'' properties of the Bernoulli shift. 

A more important issue concerns $\pi^s(D\hF_{\hX}))=D\bF_{\bX}$.
The Oseledec theorem \cite{Oseledec}
asserts that 
under mild conditions on $D\bF_{\bX}$  there exists a 
hierarchy of Lyapunov exponents 
$ \lambda_L(1) > \dots \lambda_L(N)$, Lebesgue almost-surely constant if $\hm_L$
is the SRB measure,  and a hierarchy of nested subspaces (Oseledec splitting):
$$\bbbr^N = \cV_1(\hX) \supset \cV_2(\hX) \supset \cV_N(\hX)$$
\nid depending on ${\hX}$, such that the  norm
of a perturbation $\bv \in \cV_i(\hX) 
\setminus \cV_{i+1}(\hX)$
is given by :
\beq
\|D\bF^t_{\bX}.\bv \| = C(\hX,t) e^{\lambda_L(i).t}.\|\bv\| 
\eeq

\nid where $\lim_{t \to \infty} \frac{1}{t}logC(\hX,t)=0$ almost surely,
 namely $\lambda_L(i)$
is the  exponential rate of variation of $\|\bv\|$.
Define $M(\bX,t)=\tilde{D\bF}^t_{\bX}.D\bF^t_{\bX}$ and 
$\Lambda =\lim_{t\to \infty}
M(\bX,t)^{\frac{1}{2t}}$,(the Oseledec multiplicative
ergodic theorem insures that this limit exists almost-surely and is 
a constant). Then the Lyapunov exponents are the
logarithm of the eigenvalues of $\Lambda$. $M(\bX,t)$ being
symmetric it admits an orthogonal basis 
$\left\{\bv_i(\bX,t)\right\}_{i=1}^N$ and eigenvalues $\mu_i(\bX,t)$
such that $\lambda_L(i)=\lim_{t \to \infty} \frac{1}{2t}log(\mu_{i}(\bX,t))$. 
Furthermore, each $\bv_i(\bX,t)$ converges exponentially
to a vector  $\bv_i(\hX)$ in $\bbbr^N$,
depending on $\hX$ \cite{Pollicott}. The $\bv_i(\hX)$'s constitutes therefore
a basis for the Oseledec splitting. We call them
 the {\it Oseledec modes} for the trajectory of $\hX$. They  
can be numerically obtained from the QR decomposition used in the
computation of the Lyapunov spectrum (see \cite{ER}).
It has been shown in\cite{BCK3} that  the $\lambda_L(i)$ are all negative
for finite $L$, namely all vectors in $\bbbr^N$ are asymptotically
contracted.

From this discussion, one expects a close connection between the Lyapunov
spectrum and the energy transport in the Zhang's model. In particular,
the following formula can be proved \cite{BCK3}

\beq \label{svsLyap}
\sum_{i=1}^N \lambda_L(i) = 
log(\epsilon).(1-\bar{\omega}_L) \frac{<s>_L}{<\tau>_L} = p_L log(\epsilon).<s>_L
\eeq

It relates the Lyapunov spectrum, which characterizes \textit{local} properties
of the \textit{microscopic} dynamics, to the avalanche
\textit{statistical} properties of the \textit{macroscopic} system. 
Note that the exponent $\lambda_L(i)$
gives the contraction rate in the direction $\bv_i(\hX)$ versus the {\it local
 time}.
One can also define the average contraction \textit{per avalanche},
$\chi_L(i)$,  given from eq. (\ref{tvslent}) by:

\beq \label{chiL}
\chi_L(i)= \frac{<\tau>_L}{1-\bar{\omega}_L}\lambda_L(i)=\frac{\lambda_L(i)}{p_L}
\eeq 

Then,  the sum of $\chi_L$'s, giving the average volume contraction 
\textit{per avalanche}, is related to the average avalanche size by :

\beq
\sum_{i=1}^N \chi_L(i) = log(\epsilon).<s>_L
\eeq

Note that $<s>_L$ corresponds to the total energy
transport within one avalanche and is believed to be related
to the total response function \cite{Vespignani2}. Our formula
shows that it is also equal to the  volume contraction in  phase
space produced on average by one avalanche.

\ssu{Oseledec modes.}

To each negative Lyapunov exponent $\lambda_L(i), \ i=1 \dots N$
 is associated a characteristic time 
$t_L(i)= |\lambda_L(i)|^{-1}$, the time for
of a perturbation in the Oseledec direction $i$ to vanish.
 This defines therefore a
 hierarchy:
$$t_L(1)>t_L(2)> \dots t_L(N)$$ 
\nid Note that there is no contradiction with the expected critical behaviour
in the thermodynamic limit, since as $L \to \infty$ there are an infinite number
of characteristic time scales. 

From the physical point of view a perturbation can be viewed as a small
 modification
of the initial energy landscape $\bX$. It can be  localized 
(for example one site perturbed) or spread. The  attenuation
is due to two distincts effects :

\bit

\item Propagation through the lattice.

\item Dissipation at the boundaries.
\eit 

 Note that according to the Oseledec mode under consideration
the contraction 
can  be due (on average) to the effect of one avalanche (if $t_L(i)$
is small compared to the average avalanche size),
 or to the cumulative effect of many
avalanches (if $t_L(i)$ is large). 
 The coefficient $ \chi_L(i)$ (eq. (\ref{chiL})) 
gives the average contraction per avalanche
for the $i$-th Oseledec mode. Therefore the number $\frac{1}{\chi_L(i)}$ gives
an estimate of the number of avalanches needed to have a reduction of the initial
perturbation of a ratio $\frac{1}{e}$ for this mode. Therefore a crossover
point can be estimated by :

\beq
 \chi_L(i_c) \sim 1
\eeq

We will call slow (resp. fast) modes 
the Oseledec modes corresponding to $\lambda_L(i) << \lambda_L(i_c)$
(resp. $\lambda_L(i) >> \lambda_L(i_c)$).

\sssu{Bounds on the first negative Lyapunov exponent.}

We give a bound on the first Lyapunov exponent
related to the energy dissipation at the boundaries.
Call $\Phi^{out}_j(t,\hX) \deq 1- \sum_{i=1}^N DF^t_{\bX,ij}$.
Since the energy is locally conserved, $\Phi^{out}_j(t,\hX)$ is 
the ratio of the initial energy
$X_j$ given  by the  site  $j$ to the boundary $\dL$ in $t$ time steps.
In other words, the  energy coming from $X_j$ and lost at time $t$ is
$\Phi^{out}_j(t,\hX).X_j$.  The following holds:

\bp \label{Bounds}
The largest negative Lyapunov exponent, $\lambda_L(1)$
admits the following bounds:

\beq
0> \lim_{t \rightarrow \infty} 
\frac{1}{t}log(1-\min_j(\Phi^{out}_j(t,\hX))) \geq 
\lambda_L(1)  \geq  \lim_{t \rightarrow \infty} \frac{1}{t} 
log(1 -  \max_j \Phi^{out}_j(t,\hX))
\eeq

\ep

This is interpreted as follows. As $t \to \infty$,
$\Phi^{out}_j(t,\hX)\to 1, \ \forall j, \ \forall \hX$, since, eventually,
all the initial energy coming from $\bX$  has been lost at the boundaries.  
The limit $\lim_{t \rightarrow \infty}\frac{1}{t}log(1-\Phi^{out}_j(t,\hX))$
gives the exponential rate of convergence of $\Phi^{out}_j(t,\hX)$ to $1$. 
In other words, it gives the exponential decrease for the ratio of
the initial
 energy still in the lattice at a given time. The maximal negative Lyapunov
exponent is bounded by the extremal dissipation rates.
One sees therefore that the contraction in the principal Oseledec mode is
 mainly due
to the dissipation at the boundaries. We shall see later on that $\lambda_L(1)$
is essentially related to the so-called \textit{dissipation rate}.\\

\bpr
It is easy to show that there exists a time $t_s$ depending on $E_c,\epsilon,d$
such that, whatever $\hX$ each site in the lattice has relaxed 
at least once after this time and therefore
all sites of the boundary have dissipated energy. 
At time $t$ the energy coming from a site $j$ with initial
energy $X_j$ and redistributed
into the lattice is $\sum_{i=1}^N DF^t_{\bX,ij}X_j$.
For $t\geq t_s$, $\Phi^{out}_j(t,\hX)>0$ and $\sum_{i=1}^N DF^t_{\bX,ij}$
is bounded away from $1$.
Since $D\bF^t_{\hX}$ is a matrix with positive entries:
\beq
\D{\min_j(\sum_{i=1}^N D\bF^t_{\bX,ij})} =1 -  \max_j \Phi^{out}_j(t,\hX)  \leq 
\rho(D\bF^t_\bX) 
\leq \D{\max_j(\sum_{i=1}^N DF^t_{\bX,ij})} = 1 -  \min_j\Phi^{out}_j(t,\hX) <1
\eeq

\nid where $\rho(D\bF^t_\bX)$ is the spectral radius of $D\bF^t_\bX$.
 
The largest negative Lyapunov exponent is given by :

\beq \label{deffirstlyap}
\lambda_L(1)=\lim_{t \rightarrow \infty} \frac{1}{t} log(\|DF^t_\bX\|_2)
\eeq

\nid where $\| \ \|_2$ is the $L_2$ norm.
In eq. (\ref{deffirstlyap}) the limit does not depend on $\hX$, provided $\hX$
 belongs
to the support of $\hm_L$. 
One has $\rho(D\bF^t_\bX) \leq \|DF^t_\bX\|_2$ 
and therefore:
$$\lambda_L(1) \geq  \lim_{t \rightarrow \infty} \frac{1}{t} 
log(1 -  \max_j \Phi^{out}_j(t,\hX))$$

Furthermore, all norms being equivalent in  finite dimension
eq. (\ref{deffirstlyap}) holds also for the $L_1$ norm where 
$\|DF^t_\bX\|_1=\D{\sup_{\bX}} \frac{\sum_{i,j=1}^N 
D\bF^t_{\bX,ij} |X_j|}{\sum_{j=1}^N |X_j|}$.
The $D\bF^t_{\bX,ij}$'s being positive the supremum is certainly achieved for
 positive
 $X_i$ values.
Therefore,

$$\|DF^t_\bX\|_1 = \D{\sup_{\bX}}
 \frac{\sum_{j=1}^N (1-\Phi^{out}_j(t,\hX))X_j }{\sum_{j=1}^N X_j}
= 1-\D{\inf_{\bX}}\frac{1}{\sum_{j=1}^N X_j} \sum_{j=1}^N\Phi^{out}_j(t,\hX).X_j
\leq 1-\D{\inf_{\bX}}\min_j(\Phi^{out}_j(t,\hX))$$

The limit 
$\lim_{T \rightarrow \infty}log(1-\min_j(\Phi^{out}_j(t,\hX)))$
is a constant for any $\hX$ in the support of $\hm_L$.  
Hence:
$$\lambda_L(1) \leq \lim_{t \rightarrow \infty} 
\frac{1}{t}log(1-\D{\min_j(\Phi^{out}_j(t,\hX))})$$
\epr

\sssu{Stabilizing modes}

The contraction 
in the principal Oseledec mode 
(first negative Lyapunov exponent) is mainly due
to the dissipation at the boundaries.
 On the other hand, 
it is  possible to have a large contraction in one local time step
without reaching the boundaries. Indeed, the tangent matrix $DF_{\bX}$ has the
following property, which can be checked by direct computation.

\bp \label{kernel}
Let $\Lambda = \Lambda_c(\bX) \oplus \Lambda_n(\bX)$ where $\Lambda_c(\bX) =
\left\{i \in \Lambda \ | \ X_i \geq X_c \right\}$, $\Lambda_n(\bX) =
\left\{i \in \Lambda \ | \ X_i < X_c \right\}$ and $n_c(\bX) = \#\Lambda_c(\bX)$,
then   $DF_{\bX}$ has $n_c(\bX)$ eigenvalues $\epsilon$ corresponding
to the eigenvectors 

\beq
k_i(\bX)=2.d.\be_i - \sum_{j \in \cU_i}\be_j \ ; \quad i \in \Lambda_c(\bX) 
\eeq

\nid where $\cU_i$ denotes the set of sites in $\Lambda$ at distance $1$ from
 $i$.
There are also $N-n_c(\bX)$ neutral eigenvalues associated with the eigenvectors
$\be_i, \ i \in \Lambda_n(X)$.
\ep

The eigenvectors  $k_i$ produces arbitrary large contraction as $\epsilon \to 0$.
In particular, in the original Zhang's model where $\epsilon=0$ they correspond
to \textit{kernel} modes, which have eigenvalues $0$. Note that, in this case,
the dimension of the kernel of the product tangent map $DF^t_\bX$  increases
with $t$.
 However, it 
 is strictly lower than $N$ as $t \to \infty$ \cite{BCK3}.
It is easy to check that these modes
have zero energy, except if some of the $\be_i$'s correspond
to sites neighbouring the boundary. This occurs
with small (but non zero) probability. These modes act as directions
where a single  local time step is sufficient to reduce the initial perturbation
by a factor $\epsilon$, with
 small variation of the total energy on average.
 They dynamically correspond to directions
transverse to the attractor, and their contraction corresponds to a fast
 convergence
onto the attractor. For this reason we call them {\it stabilizing modes}.
In the Lyapunov spectrum they can be identified because the corresponding
Lyapunov exponents go to $-\infty$ as $\epsilon \to 0$ while
the other part of the spectrum weakly depends on $\epsilon$ (see Fig \ref{Fkern}).
 
\bef
 \bc
 \begin{minipage}{10cm}
  \epsfxsize=10cm
 \epsfysize=6cm
 \epsffile{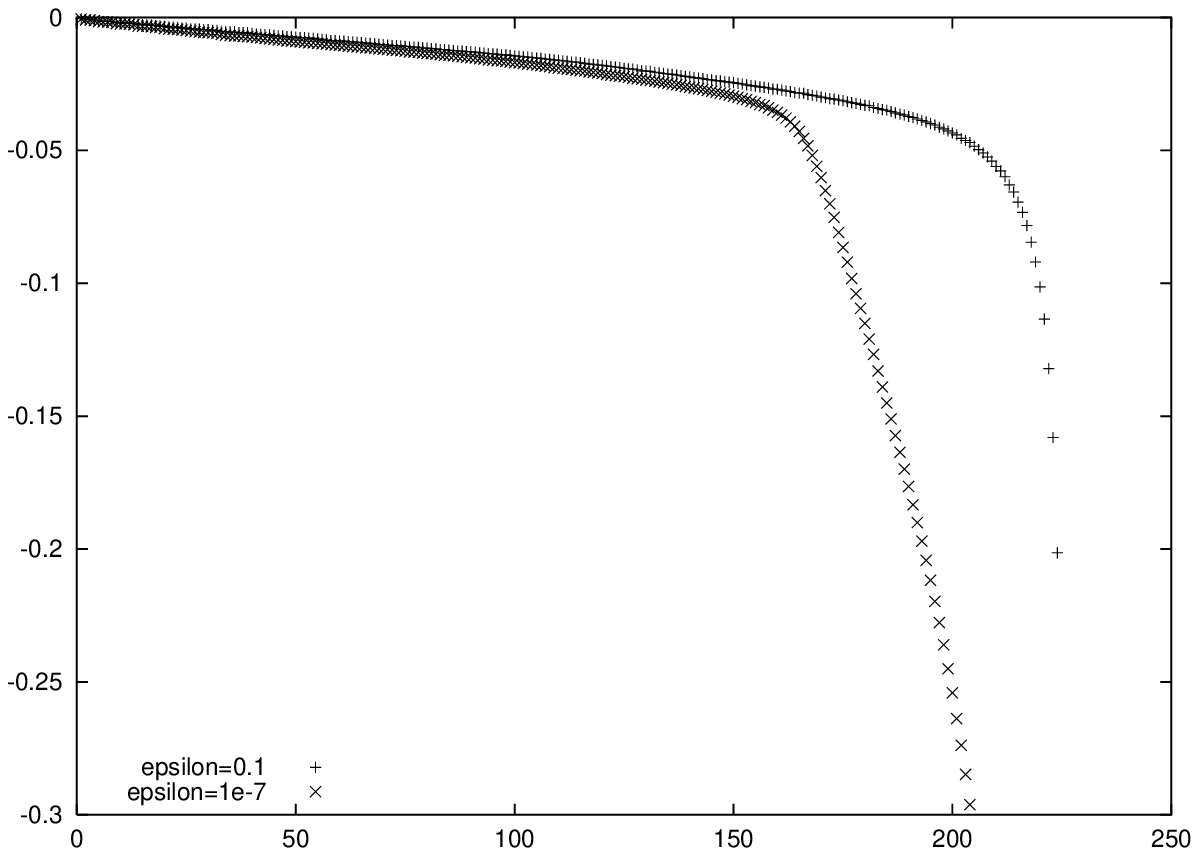}
\end{minipage}
\hspace{1cm}
\vspace{0.3cm}

\caption{Lyapunov spectrum for $\epsilon=0.1$ and 
$\epsilon=10^{-7}$, $L=15$, $E_c=2.2$.
}\label{Fkern} 
 \ec
 \enf

\sssu{The transport operator.}

It is usually not possible to give an explicit formula
for the whole Lyapunov spectrum, except in some specific
cases \cite{Newmann}. We propose here a mean-field
ansatz which gives good results for the slowest modes,
and has a nice interpretation in terms of random walk.
It is based on the following observations.

The Lyapunov exponents are the eigenvalues of the
matrix $\Lambda=E_L\left[\Lambda \right]=
E_L\left[\lim_{t \to \infty} 
\left(\tilde{D}\bF^t_{\bX}.D\bF^t_{\bX} \right)^{\frac{1}{2t}}\right]$.
Since the matrix $\tilde{D}\bF^t_{\bX}.D\bF^t_{\bX}$ is bounded 
in $L_2$ norm, $\forall \bX, \forall t$, one has from the Lebesgue
theorem : $\Lambda=
\lim_{t \to \infty} 
E_L\left[\left(\tilde{D}\bF^t_{\bX}.D\bF^t_{\bX} \right)^{\frac{1}{2t}}\right]$.
On the other hand, the matrix 

\beq
\cL(t)=E_L\left[D\bF^t_{\bX}\right]
\eeq 

\nid characterizes the (ensemble) average energy transport in $t$ time steps.
However, the connection between $\cL(t)$ and $\Lambda$  is 
loose.

Were the transport be normal, namely were  $D\bF_{\bX}$ be independent
of $\bX$ and of the form $D\bF_{\bX}=I+\gamma.\Delta$, where $\gamma$
is some constant, then were $\cL(t)$ be equal to $(I+\gamma.\Delta)^t$.
In this case, $\cL(t)$ would be constant and symmetric. Then 
$E_L\left[\left(\tilde{D}\bF^t_{\bX}.D\bF^t_{\bX} \right)^{\frac{1}{2t}}\right]
= I+\gamma.\Delta$
and $\Lambda=I+\gamma.\Delta$. In this case the Lyapunov exponents would
be the eigenvalues of a  one-step transport operator $\cL=I+\gamma.\Delta$
(Fourier modes).

More generally, the (naive) hope would be that of finding an effective
transport operator $\cL$ such that $\cL(t)=\cL^t$ 
and whose singular values (or eigenvalues if $\cL$ is self-adjoined)
would give the Lyapunov spectrum. There is  however a priori no hope
for finding such an operator, in general. Note in particular
that the assumption of independence of the matrices 
$D\bF_{\bX(t)}$, first step towards a mean-field approach,
is not a sufficient condition.
Since, in this case $E_L\left[D\bF^t_{\bX}\right]=
E_L\left[D\bF_{\bX}\right]^t$, 
one is lead to propose $\cL=E_L\left[D\bF_{\bX}\right]$
as a one-step operator. However, one needs further conditions
to insure that the singular values of $\cL$ give the Lyapunov exponents
(see for example \cite{Newmann,Eckmann}).
 It appears nevertheless that
in the Zhang's model an effective transport operator can be found
from a mean-field treatment
which well approximates the \textit{slowest modes}.\\

The first obstacle towards a mean-field approach lies in the independence
 assumption. 
The matrix $\cL(t)$ is a sum of time correlations terms of the form 
$E_L\left[S(\bX(t_{r-1}))S(\bX(t_{r-2})) \dots S(\bX(t_0)) \right]$
whose entry $(i,j)$ writes $\sum_{i_1, \dots, i_{r-1}}
\Delta_{i,i_{r-1}}\dots
\Delta_{i_2,i_1}\Delta_{i_1,j}
Prob\left[Z_{i_{r-1}}(\bX(t_{r-1}))=1,\dots,Z_{i_1}
(\bX(t_1))=1,Z_j(\bX(t_0))=1\right]$.
Clearly, the non-vanishing terms in this sum are those corresponding to a path
from $j$ to $i$ where each intermediate site has been active at least once
with a non zero probability.
A simple glance at this formula shows that 
\textit{a priori all time correlation functions of the joint probability of 
active
 sites},
$Prob\left[Z_{i_{r-1}}(\bX(t_{r-1}))=1,\dots,Z_{i_1}(\bX(t_1))=1,Z_j(\bX(t_0))=1\right]$
have to be considered. 

However,   Zhang's model,  as a hyperbolic
dynamical system, has exponential correlation decay (for finite $L$).
The largest correlation decay rate is given by the entropy
$\bar{\omega}_L.log(N)$. This decay rate is quite faster than the characteristic
times related to the slow modes (for example the correlation decay rate 
of a site with itself is about $-0.025$ for $E_c=2.2,\epsilon=0.1,L=20$,
corresponding roughly to the $320$-th exponent in the spectrum,
while the slowest Lyapunov exponent value is $-0.000209871$).
More generally,  we show 
in the last section, that $\lambda_L(1) \sim \frac{\bar{\omega}_L}{L^d}$
to be compared to the decay rate $\bar{\omega}_L.log(N)$.
On the other hand, for the slow modes,
a small
perturbation has essentially no variation during one step of an avalanche.
In other words, the slowest Oseledec modes are not sensitive to the fast 
changes (one local time step) of the individual
matrices $D\bF^t_{\bX}$ (resp. the fluctuations of the $Z_j(\bX(t))$'s)
but, rather, to the average variations on the characteristic 
time scale
$t_L(i)=\frac{1}{\lambda_L(i)}$,
 which is quite longer than a local time step.
 This suggests that one should consider the projection
of the matrices $D\bF_{\bX(t)}$'s on the slow Oseledec space 
as independents. This leads  
to propose $E_L\left[D\bF_{\bX}\right]=I+\alpha.\Delta.\rho_L.I$
as a one-step transport operator. Note that we obtain the same
result by assuming that the $Z_{i}(\bX(t))$'s are \textit{independent}.
Indeed,
in this case $Prob\left[Z_{i_{r-1}}(\bX(t_{r-1}))=1,\dots,Z_{i_1}(\bX(t_1))=1,
Z_j(\bX(t_0))=1\right]
= \rho_L(i_{r-1})\dots\rho_L(i_1)\rho_L(j)$. Then
$\cL(t)=\sum_{k=1}^t C_t^k((\Delta.\rho_L.I)^k=(I+\alpha\Delta.\rho_L.I)^t$.\\

This approximation gives correct  results .... provided that one multiplies
the density of active sites by $2$ !!! 
This approximation neglects indeed an important effect.
Provided $E_c > \frac{\epsilon}{1-\epsilon}$, a site \textit{cannot
relax two successive time steps} \cite{BCK3}, and therefore it relaxes
at most only half of the time during one avalanche.
This means in particular that the random variables $Z_i(\bX(t)),Z_i(\bX(t+1))$ 
are 
\textit{not independent} and that the probability that one site relaxes at time
 $t+1$
depends on what happened at time $t$. Besides,
two neighbouring sites cannot be simultaneously active.
In a certain sense, the lattice is ``blinking'' :
 during one avalanche all active sites are
at pairwise distance \cite{BCK3}. This therefore introduces  strong correlations
between $\bZ(\bX(t))$ and $\bZ(\bX(t+1))$.

 One can, however, circumvent the problem
by reparametrizing the time and considering the evolution of the 
process every \textit{two times steps}.  Equivalently, one replaces 
the stochastic process $\left\{\bZ(\bX(t))\right\}_{t=1}^{+\infty}$ by
a new process 
$\left\{\bY(t')\right\}_{t'=1}^{+\infty}=\left\{\bZ(\bX(t)),\bZ(\bX(t+1))
\right\}_{t=1}^{+\infty}$
whose components $Y_k(t)$ take values in $\left\{0,1\right\}^2$,
 where the event $(1,1)$
has zero probability and where $t'=\frac{t}{2}$. One can then 
encode the $Y_k(t')$ values by $0 \rightarrow (0,0)$ (no relaxation at
times $t,t+1$)
and $1 \rightarrow (0,1),(1,0)$ (relaxation at time $t$ or at time $t+1$).
This leads to the definition of  a new ``density of active sites''
$\rho'_L(i)=Prob\left[ Y_i(t')=1\right]=Prob\left[Z_i(\bX(t))=1 \ or \
Z_i(\bX(t+1))=1 \right]$. Since the events $\left\{Z_i(\bX(t)=1\right\}$
and $\left\{Z_i(\bX(t+1)=1\right\}$ are disjoints, we have:
$\rho'_L(i)=Prob\left[\bZ(\bX(t))=1\right] +
Prob\left[\bZ(\bX(t+1))=1 \right]=2.\rho_L(i)$.
Assuming now that the $Y_k(t')$'s are independent and considering
$\rho'_L(i)$ as the effective density of active sites,
one obtains
 an effective transport operator:

\beq \label{trans}
\cL=I+2\alpha\Delta.\rho_L.I
\eeq

Calling $\gamma_i$ the singular values of $\cL$, our
mean-field ansatz suggests that the slowest Lyapunov modes
are given by :

\beq 
\lambda_L(i)=log(\gamma_i)
\eeq

Note that this operator  is self-adjoined
for the metric $\rho_L.I$ and that the corresponding
matrix can be made  symmetric by the variable change 
$\rho_L^{-\frac{1}{2}}.I$.

To check the validity of this ansatz, we first
computed the density of active sites on a $20 \times 20$ lattice
and found numerically the $\gamma_i$'s from these data
\footnote{We weren't able to go beyond $L=20$ in the Lyapunov spectrum 
computation.
We used a version of the Eckmann-Ruelle algorithm \cite{ER} revisited from
Von Bremmen et al. \cite{VonBremmen}. Nevertheless, we needed two weeks of 
computation
on a Pentium II 300 for the case $L=20$, with a relative accuracy
of $10^{-3}$.}.
 At the
same time we computed the Lyapunov spectrum. A plot of the two curves
is drawn fig. \ref{Fdiflyap}. One finds 
a very good agreement over a large part of the spectrum,
and the discrepancy increases towards small times scale,
as expected.

\bef 
 \bc
 \begin{minipage}{10cm}
  \epsfxsize=10cm
 \epsfysize=6cm
 \epsffile{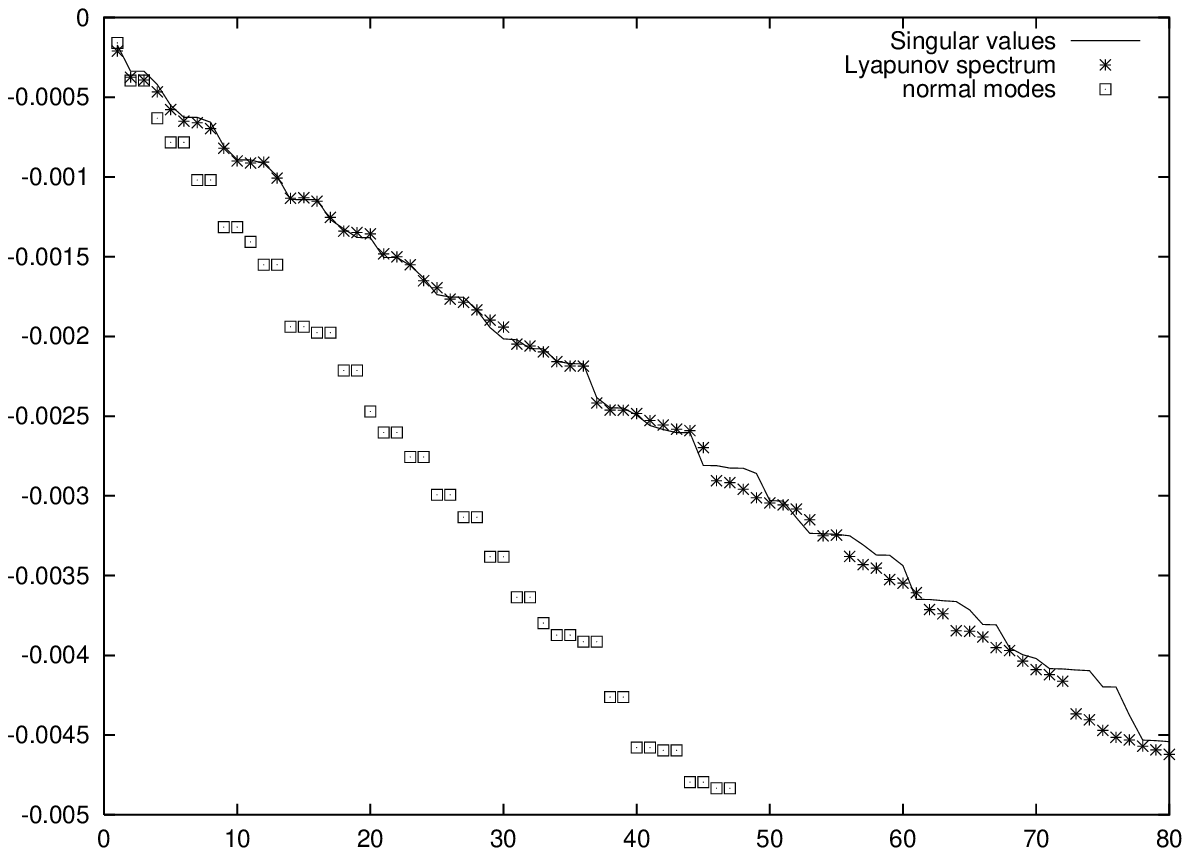}
\end{minipage}
\hspace{1cm}
 \begin{minipage}{10cm}
  \epsfxsize=10cm
 \epsfysize=6cm
 \epsffile{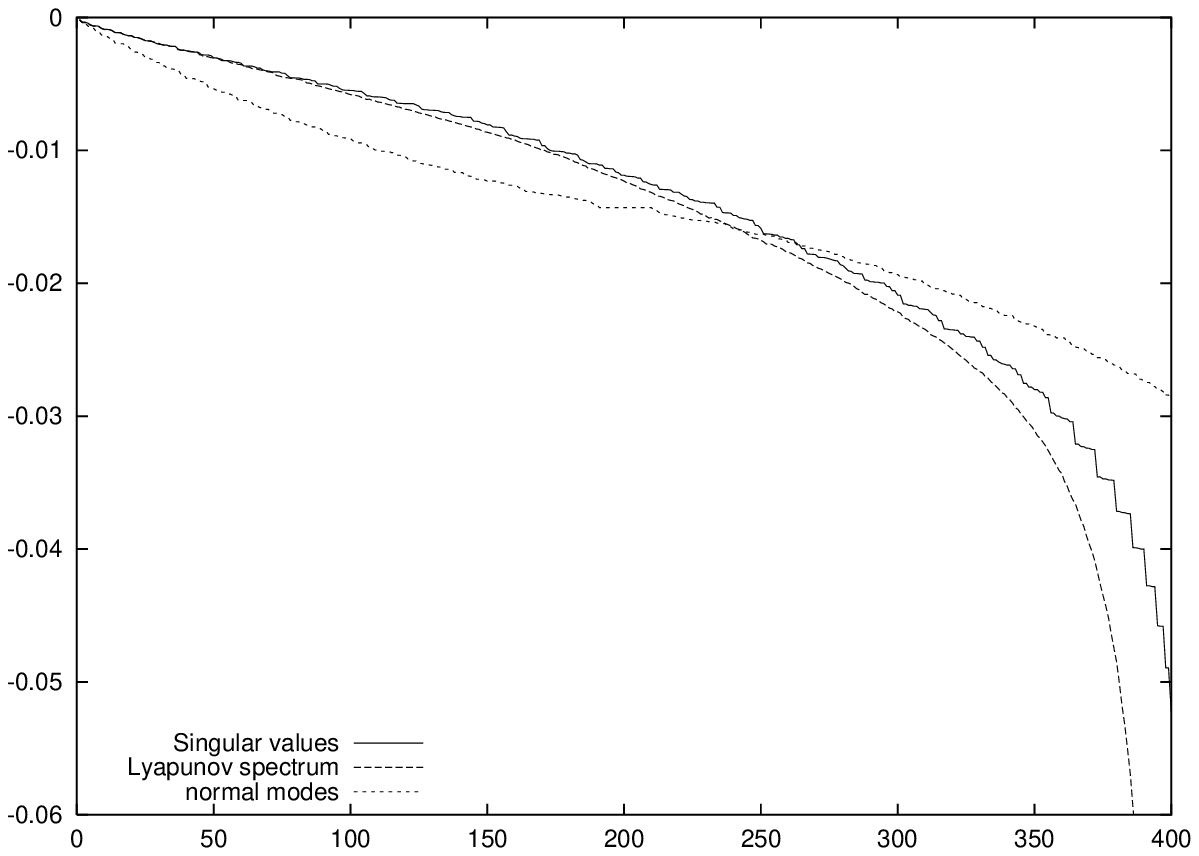}
\end{minipage}
\vspace{0.3cm}

\caption{Lyapunov spectrum, logarithm of the $\cL$ singular values, and
normal diffusion modes
for $E_c=2.2$,$\epsilon=0.1$,$L=20$. Fig \ref{Fdiflyap}a : 80 first modes;
 Fig \ref{Fdiflyap}b : Full spectra drawn with lines in order to better
see the shape.}\label{Fdiflyap}
 \ec
 \enf

\sssu{The role of the spatial variations in the density of active sites.}

It is usually assumed in the SOC literature, when dealing
with the model's spatial properties,
that only the density of active sites, and more
precisely, its lattice average $\rho_L^{av}$,  has to be taken into account. 
One therefore neglects the spatial dependence of $\rho_L$. 
In our approach this would 
lead to an effective transport operator $I+2.\alpha\rho_L^{av}\Delta$,
corresponding to normal transport.
 In this case, the slowest Oseledec modes would  simply be the 
eigenmodes of the Laplacian with zero boundaries conditions on $\dL$,
and the Lyapunov exponents would then correspond to the normal diffusion modes:

\beq \label{normaldiflyap}
\lambda_L(i)  = 
log(|1+2.\alpha.\rho^{av}_L.\sum_{k=1}^d(cos(\frac{\pi.n_k}{L+1})-1)|)
\eeq

\nid where the Laplacian modes are parametrized by the quantum numbers 
$\bn=(n_k), k=1 \dots d$,
 sorted such that the corresponding eigenvalues are decreasing (and $i$ refers 
to the
 placement of
the exponent in this sequence). 
In fig. (\ref{Fdiflyap}), we also plot  the diffusion eigenvalues of eq. 
(\ref{normaldiflyap}).
The computed Lyapunov exponents are different from these values
except for the largest ones. 
This approximation is therefore too crude and gives a wrong
spectrum. 
 Note in particular that the {\it shape}
of the spectrum differs, namely the discrepancy cannot
be corrected by a mere multiplication of $\rho^{av}_L$
by some factor. Since the Lyapunov exponents contain
all the relevant informations about the dynamics at stationarity, our conclusion
is that the non-homogeneity of $\rho_L(i)$ plays a key role in computing
dynamical quantities, and implies unfortunately that the zero-th order
``mean-field'' approaches,
which  approximate the density of active sites as a constant lead to 
incorrect estimates for finite size when dealing
with intermediate time scales. On the other hand,
this  should lead to correct results when 
dealing with the longest time scales,  since the first
modes are well approximated by a transport operator
where $\rho_L(i)$
is considered as uniform.

In the litterature one  often meets an (apparent) contradiction
(see for example the original paper form Zhang \cite{Zhang}
and subsequent analysis by Pietronero et al. \cite{Pietronero}).
One assumes the transfer of energy on large time scales
to be normal diffusion, while in the same time an
anomalous diffusion exponent $z \neq 2$ is computed.
 It was certainly clear in the spirit of these authors
that one has to distinguish the average transport on many
avalanches (long time scales) from the average transport within
one avalanche (characterized by $z$). Our results on the Lyapunov spectrum 
makes this distinction quantitative.
We show that the transport on the longest 
time scales is  normal with a good approximation, while  the
average transport within \textit{one} avalanche, namely on time scales
corresponding roughly to the crossover point $\lambda_L(i_c)$, is
clearly anomalous.

\sssu{The random walk picture.}

The operator $\cL$ is the Laplace-Beltrami operator associated
to a random diffusion in a ``medium'' or a landscape
\textit{that is not flat}, corresponding
to the metric $g=\rho_L.I$, where $I$ is the identity matrix on $\bbbr^N$.
It has a nice interpretation in the
so-called random walk picture 
\footnote{B.C. is very grateful to P. Grassberger and D. Dhar
for illuminating discussions on this point in Trieste.}.
Assume for a moment that the energy
of a site is composed by  (undivisible) energy quanta $\eta$
that can be made arbitrarily small (this is way to ``map'' the Zhang's model
to a sandpile).
Assume that we are 
in the stationary regim, and that at the initial time
we drop a grain in some place and  study its motion. At
each time step where it is involved in a relaxation process
this grain  makes a jump at random in one of the $2.d$ directions in the lattice.
From this point of view, the stochastic dynamics of the grain is driven by the
 underlying
dynamics of eq.(\ref{relaxation}). 
\textit{If we assume that the evolution is Markovian}, 
the probability of jumping from $i$ to some nearest neighbour $j$
depends only on the state of $i$ at time $t$ and is given
by a transition rate $W_{ij}=\alpha.\rho_L(i)$,
while the  probability of staying at the same place is $1-\rho_L(i)(1-\epsilon)$ 
(remember that only an amount $\alpha=\frac{1-\epsilon}{2d}$ of the energy is transferred
to the neighbours when a site relaxes).
From this consideration, one obtains the equation for the probability $P(i,t)$
that a grain is at place $i$ at time $t$, before it leaves the lattice
$P(.,t+1)=[I + \alpha.\Delta\rho_L].P(.,t)=[I + \alpha.\Delta\rho_L]^t .P(.,1)$,
and one recovers the operator obtained
above when assuming that the $Z_i(t)$'s were independent.
Indeed, the independence assumption of the $Z_i(t)$'s
is \textit{equivalent to the Markovian assumption in the
random walk picture}. The probability of jumping for a grain
at time $t$ depends \textit{a priori} on its whole past via a Chapman-Kolmogorov
equation whose transfert matrix is a sum of terms
containing conditional probabilities 

$$Prob\left[Z_{i_{r-1}}(\bX(t_{r-1}))=1|
Z_{i_{r-2}}(\bX(t_{r-2}))=1,\dots,Z_{i_1}(\bX(t_1))=1,Z_j(\bX(t_0))=1\right]=$$
$$\frac{Prob\left[Z_{i_{r-1}}(\bX(t_{r-1}))=1,
Z_{i_{r-2}}(\bX(t_{r-2}))=1,\dots,Z_{i_1}(\bX(t_1))=1,Z_j(\bX(t_0))=1\right]}
{Prob\left[Z_{i_{r-2}}(\bX(t_{r-2}))=1,\dots,Z_{i_1}(\bX(t_1))=1,
Z_j(\bX(t_0))=1\right]}=$$
$$Prob\left[Z_{i_{r-1}}(\bX(t_{r-1}))=1\right]$$

\nid where the last equality holds when the $Z_k(t)$'s are independent.
In this case, $W_{ij}=\alpha.Prob\left[Z_{i}(\bX(t))=1\right]=\alpha\rho_L(i)$
for the $j$ nearest neighbours of $i$.

However, we saw above that the process is not strictly Markovian
since a jump from a given site cannot take place at two successive
time steps. In other words, the probability
of a jump $i \to j$
depends on the state of $i$ at time $t$
\textit{and} at time $t-1$.
The system has some memory (at least two-time-steps).
 However, defining the random variables
$Y_k(t)$'s, as above, and assuming them to be independent
amounts to render the random walk Markovian by a suitable reparametrization
of the process, and gives a transfert equation 

\beq \label{difP}
P(.,t+1)=[I + 2.\alpha.\Delta\rho_L].P(.,t)= \cL^t.P(.,1)
\eeq

Therefore, the operator $\cL$ characterizing the decay of a small
perturbation can also be interpreted as the transfert matrix
of a random walk in a medium where the diffusion rate
depends on the location.

\sssu{Density of active sites and average energy flowing towards the boundaries.}

Eq. (\ref{difP}) characterizes the energy transport in the lattice,
but does not
take into account the source term (addition of a grain) required
to reach stationarity.  Indeed, 
each times 
a grain exits the lattice, one must add another grain 
at a random place $i$,
with probability $\bar{\omega}_L(i)$ ; this  is a source term.
 Call $\cP_L$ the equilibrium state
of the random walk and $\cV(\partial\lambda)$ 
 the set of sites at distance one from the boundary. 
The probability for a grain to exit is 
 $2\alpha \sum_{j\in \cV(\partial\lambda)}\rho_L(i).\cP_L(i)$. 
It is obviously proportional to the
outgoing energy flux, which is,  at stationarity,
 equal to the incoming flux (resp. the probability of adding a grain in the lattice),
 namely
\footnote{The cautious reader has noticed that this equation is not dimensionally
correct, since no energy term appears on the l.h.s.. One should indeed multiply
by $\delta$, the input energy quantum, which is set to one throughout this
paper.}:

\beq \label{balance}
\bar{\omega}_L=2\alpha\sum_{j\in \cV(\partial\lambda)}\rho_L(j).\bar{X}_L(j)
\eeq

The  complete equation for the energy at stationarity is :

\beq\label{Estat}
2\alpha.\Delta[\rho_L.\bar{\bX}_L]+\bar{\omega}_L(i)=0
\eeq

\nid with zero boundaries conditions and with the constraint (\ref{balance}).

 In this equation one distinguishes a {\it local} transport
term, and a source term which depends on a {\it global} constraint.
When the excitation is uniform eq. (\ref{Estat}) reduces to 
$\Delta[\rho_L.\bar{\bX}_L]+\frac{\bar{\omega}_L}{2\alpha.L^d}=0$.\\

The difficulty in solving this equation is that it deals with the product
$\rho_L.\bar{\bX}_L$.  On the other hand, it is known
in the literature that $\bar{\bX}_L$ converges to a uniform
energy distribution over the lattice as $L \to  \infty$ \cite{Grassberger1}.
Assume now that we can
write  $\bar{\bX}_L$ as:

\beq
\bar{\bX}_L = \bar{\bX}_0+ f(L)
\eeq

\nid where $\|f(L)\|$ goes to zero as $L \to  \infty$ and where $\bar{\bX}_0$
is spatially uniform, i.e. $\bar{\bX}_0(i)=const=\bar{x}_L$.
At the zero-th order, one obtains for
$\rho_L$ the following equation:

\beq\label{Estatpremierordre}
\Delta\rho_L+\frac{\bar{\omega}_L}{2.\alpha L^d \bar{x}_L}=0
\eeq

\nid where $L^d \bar{x}_L= E_{tot}$, the average total energy in the lattice.
 The solution of this equation can be easily
found by decomposition on the eigenmodes of the Laplacian. 
 The general solution   is:

\beq \label{rhoL}
\rho_L(\bx) = \sum_{\bn} A_{\bn} \prod_{i=1}^dsin(k_i.x_i)
\eeq

\nid where  $\bn=(n_1,\dots, n_d)$ is the set of quantum numbers
parametrizing the eigenmodes of the discrete Laplace operator,
$s_{\bn}=2(\sum_{i=1}^dcos(k_i)-d)$ is the corresponding  eigenvalue
with $k_i=\frac{n_i\pi}{L+1}$,

$$A_{\bn}=-\frac{2^{d-1}.\bar{\omega}_L}{\alpha E_{tot}(L+1)^d}
\frac{\prod_{i=1}^dC_{n_i}}
{s_{\bn}}$$

\nid and
$$C_{n_i}=\sum_{x=1}^L sin(k_i.x)
=(-1)^{m_i}\frac{sin(\frac{n_i\pi L}{2(L+1)})}{sin(\frac{n_i\pi}{2(L+1)})}$$
 
\nid where $n_i=2m_i+1$.
Surprisingly, this already gives  quite a good approximation for $\rho_L$,
which becomes better and better as $L$ increases (see fig. \ref{FrhoL} and fig. 
\ref{FrhoLmoy}
in the next section.).

\bef
 \bc
 \begin{minipage}{10cm}
  \epsfxsize=10cm
 \epsfysize=6cm
 \epsffile{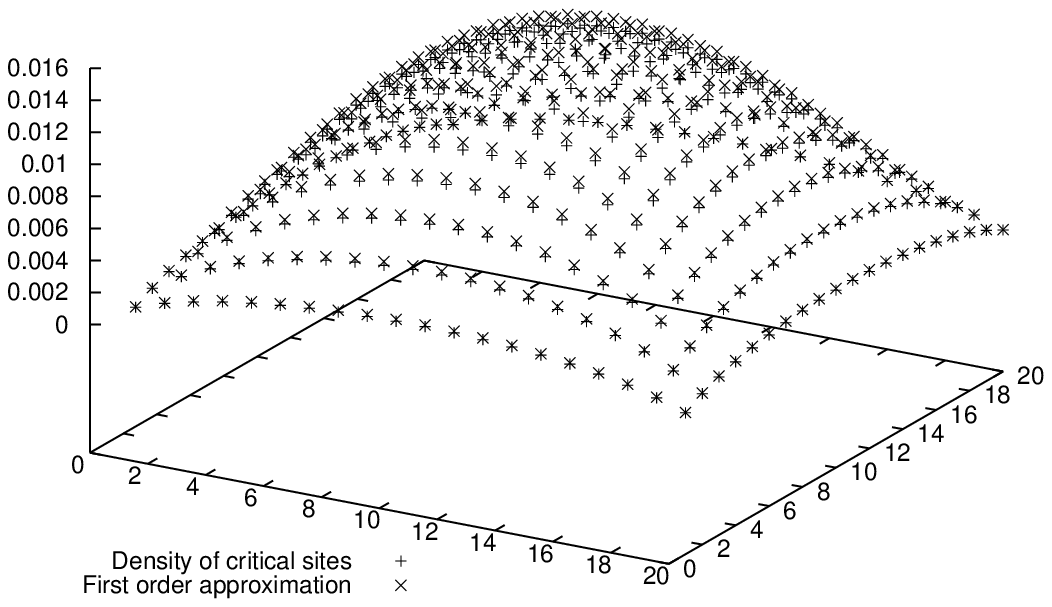}
\end{minipage}
\vspace{0.3cm}

\caption{Plot of the density of active sites and solution of eq. 
(\ref{Estatpremierordre})
for $E_c=2.2$,$\epsilon=0.1$,$L=20$.} \label{FrhoL}
 \ec
 \enf

Away from the boundaries, one expects rotational invariance for $\rho_L(\bx)$.
This can be checked by expanding the function $\sin$ near to $x_i=\frac{L}{2},
i=1 \dots d$ up to  third order. One obtains the well known paraboloid form
\cite{Grassberger}
$\rho_L(\bx) \sim K_0 - K_1\sum_{i=1}^dx_i^2$, where the constants $K_0,K_1$
can be easily deduced from eq. (\ref{Estatpremierordre}).

One also obtains the average density of active sites,
$\rho_L^{av} = \frac{1}{L^d}\sum_{i=1}^N\rho_L(i)$: 

\beq\label{rhopremierordre}
\rho_L^{av} =-\frac{2^{d-1}.\bar{\omega}_L}{L^d(L+1)^d\alpha E_{tot}}
\sum_{\bn}\frac{\prod_{i=1}^dC^2_{n_i}}
{s_{\bn}}
\eeq

\nid which is expected to hold for sufficiently large $L$. We give a plot
 fig.
\ref{FrhoLmoy} where it clearly appears that this formula gives already a quite good
 estimate
for $L=15$.

\bef
 \bc
 \begin{minipage}{10cm}
  \epsfxsize=10cm
 \epsfysize=6cm
 \epsffile{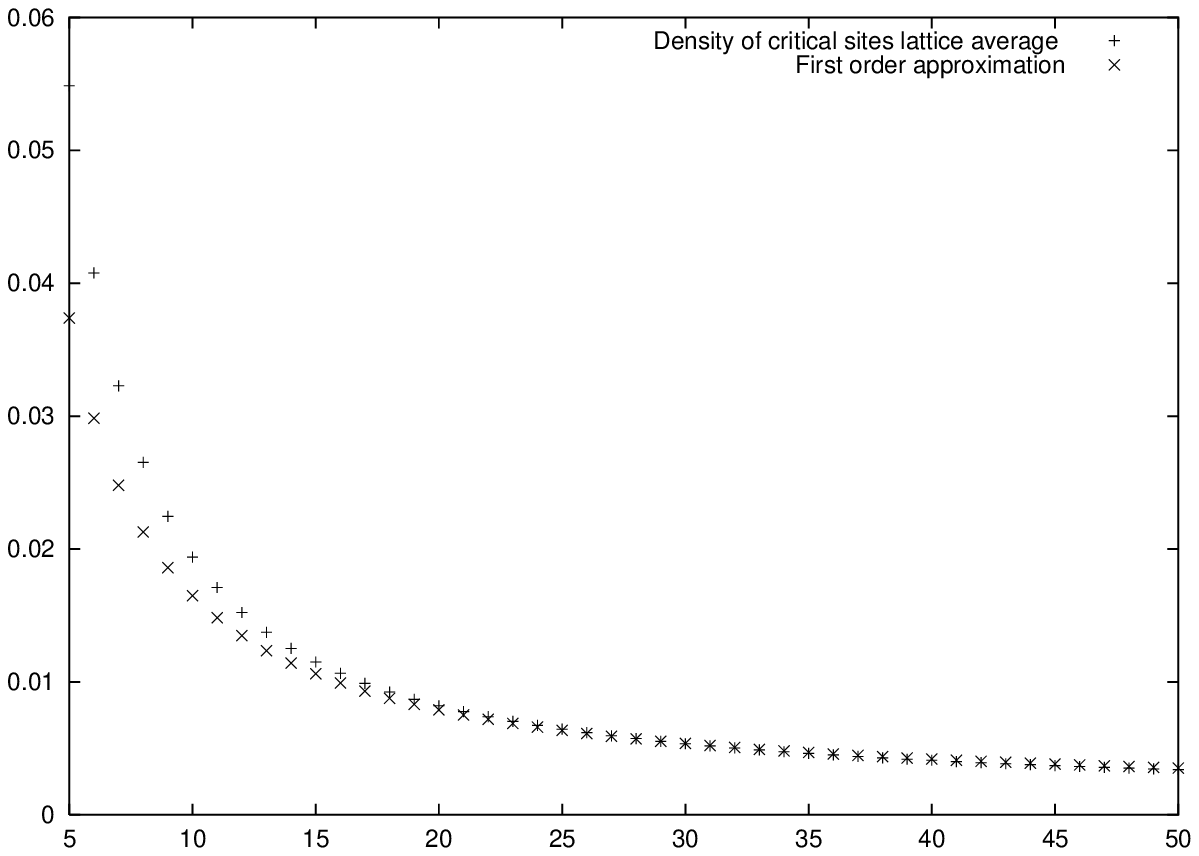}
\end{minipage}
\vspace{0.3cm}

\caption{Plot of $\rho_L^{av}$  and solution of eq. 
(\ref{rhopremierordre}) versus L,
for $E_c=2.2$,$\epsilon=0.1$.} \label{FrhoLmoy}
 \ec
 \enf

\su{Scaling properties of the Lyapunov spectrum.}

Zhang's model, as a hyperbolic dynamical system, cannot exhibit
a critical behaviour for finite size, since it has exponential
correlation decay \footnote{The exponential correlation decay
is a general property of hyperbolic systems but in presence
of singularities one can also observe a polynomial correlation decay
and weak initial condition senstivity \cite{BCK1}.}. 
However, since a critical behaviour is  conjectured in the
thermodynamic limit, one expects that hyperbolicity
is lost as $L \to \infty$, namely some of the Lyapunov exponents
go to zero. It is therefore of crucial importance to know the behaviour
of the Lyapunov exponents as $L \to \infty$. In this section, we
first discuss the time scale separation between activation rate and dissipation
rate,
which is believed to be a fundamental ingredient to have SOC, and its links
to the Lyapunov exponents. 
We then show that using a Finite Size
Scaling ansatz provides a scaling exponent
from which the scaling of some SOC observables can be obtained.

\ssu{Time scale separation.}

Equation (\ref{rhopremierordre}) can be written as:

\beq
\rho_L^{av} =-\frac{2^{d-1}.\bar{\omega}_L}{L^d \alpha E_{tot}}.\gamma_L
\eeq

\nid where $(L+1)^d.\gamma_L=\sum_{\bn}\frac{\prod_{i=1}^dC^2_{n_i}}
{s_{\bn}}$. Let us estimate the scaling
of this sum as $L \to \infty$. First, set  $d=1$ and fix $\alpha >0$ arbitrarily 
small.
The sum over $n=n_1$ can be split into a part such
that $n < (L+1)\alpha$ and another part where $n \geq (L+1)\alpha$.
In the first sum, $C_{n} \sim \frac{2(L+1)}{n\pi}$,
$s_{\bn} \sim -\frac{(\pi.n)^2}{(L+1)^2}$, while
the second sum is smaller than  $(L+1)C(\alpha)$, where $C(\alpha)$
is bounded for $\alpha >0$. Therefore $\gamma_L \sim
(L+1)^3.S$, where $S \sim \sum_{n < (L+1)\alpha }\frac{1}{n^4}$
stays bounded as $L \to \infty$. Then $\gamma_L \sim
(L+1)^3$. This argument can be generalized for any d
by splitting the sum over $\bn=(n_1,n_2, \dots n_d)$ into
sums where $k$  indexes are smaller that $\alpha(L+1)$,
$k$ going from $0$ to $d$. It is easy to see that the main
contribution is due to the terms such that  $d$
indexes are $< \alpha(L+1)$, giving a leading contribution
$\sum (L+1)^{2d+2}$ and $\gamma_L \sim L^{d+2}$. 
We therefore conclude  that $\rho_L^{av}$
 scales like :

\beq \label{rholav}
\rho_L^{av} \sim \frac{\bar{\omega}_L}{E_{tot}}\frac{(L+1)^{d+2}}{L^d}
 \sim\frac{\bar{\omega}_L.L^{2}}{E_{tot}} = 
\frac{\bar{\omega}_L.L^{2}}{L^d\bar{X}_0}
\eeq

\nid

Set $h=\frac{\bar{\omega}_L}{L^d}$ for the driving rate
and call $e=\frac{h}{\rho_L^{av}}=\bar{x}_L.L^{-2}$.
One obtains the energy conservation equation : $h=\rho_L^{av}e$
and therefore $e$ is the energy dissipated per active site
and per unit time. It corresponds to the \textit{dissipation rate} 
introduced by Vespignani et al.\cite{Vespignani2}. Since
$0 <\bar{x}_L<E_c, \forall L$, $\bar{x}_L$ plays no role
in the asymptotic scalings in $L$ and therefore
$e \sim L^{-2}$, as already anticipated by a mean-field approach
in \cite{Vespignani2}.

The average value of observables like size, duration, etc ...
 is known to diverge
with a power law scaling $\left< x \right>_L \sim L^{\gamma_x}$.
Therefore $\frac{1}{\left<\tau\right>_L} \to 0$ like $L^{-\gamma_\tau}$
as $L \to \infty$ where $\gamma_{\tau} > 1$
\cite{Jensen}. Since $0 \leq \bar{\omega}_L \leq 1$ (see (\ref{bom})),
eq. (\ref{plomega}) 
implies that $p_L=f(\frac{1}{\left<\tau\right>_L}) =
 \frac{a_1}{\left<\tau\right>_L} - \frac{a_2}{\left<\tau\right>^2_L}
+O(\frac{1}{\left<\tau\right>^3_L})$. This is in particular 
clear for $E_c < 1$, since $p_L= \bar{\omega}_L$, which implies
$p_L=\frac{1}{1+\left<\tau\right>_L}$ and therefore $a_1=1, a_2=1$.
For general $E_c$ using this form gives, from eq. (\ref{plomega}) $a_1=1$
and :

\beq \label{scalingpL}
p_L \sim \frac{1}{\left<\tau\right>_L}-\frac{a_2}{\left<\tau\right>^2_L}
\eeq

\nid and

\beq \label{scalingomegal}
\bar{\omega}_L \sim \frac{a_2}{\left<\tau\right>_L}
\eeq

\nid as $L \to \infty$. It follows therefore that :

\beq \label{scalingrholav}
\rho_L^{av} \sim L^{-\gamma_{\tau}+2-d}
\eeq

We have therefore shown that :

\beq \label{triplelim}
h \to 0, e \to 0, \rho_L^{av}=\frac{h}{e} \to 0, \ \mbox{as} \ L \to \infty 
\eeq

In \cite{Vespignani2} Vespignani et al.
discussed the necessity of this triple limit  in order to have SOC.
In their analysis  the activation and dissipation
rate where however free parameters (tunable ``by hand'').
 In Zhang's model, $h$ and $e$ are not free, since they are fully
determined by the dynamics. Therefore, we have shown
 that the three limits discussed by Vespignani et al.\cite{Vespignani2}
are indeed achieved,  without external fine tuning of some parameter,
in  Zhang's model, by the simple constraints one imposes
on the dynamics (adiabatic driving). 

 From (\ref{scalingomegal}) we have that the positive
Lyapunov exponent (resp. the entropy) $\bar{\omega}_L.log(N) \to 0$
in the thermodynamic limit. On the other hand, 
the first negative Lyapunov exponent is given with a
good accuracy by the normal diffusion operator $1+2.\rho_L^{av}\Delta$ (see
fig.\ref{Fdiflyap}), which implies that $\lambda_L(1) \sim \rho_L^{av}L^{-2}$.
Another way of arguing is to note that from theorem 1, $\lambda_L(1)$
scales like the average ratio of energy dissipated by one site.
From the local conservation of energy,
 $\lambda_L(1).\bar{x}_L \sim h$, then  $\lambda_L(1) \sim \rho_L^{av}L^{-2}$. Therefore,  $\lambda_L(1) \to 0$
in the thermodynamic limit. 
Actually, the Finite-Size Scaling analysis of the next section suggests
that a large number of negative Lyapunov exponents also
go to zero as $L \to \infty$. But the double limit $\lambda_L(0) \to 0,
\lambda_L(1) \to 0$ already show that the \textit{hyperbolicity is lost
in the thermodynamic limit}. Note however that these
two exponents are not indepedent since local conservation of energy
imposes $\lambda_L(1).\bar{x}_L \sim h$ which implies
$\frac{\lambda_L(1)}{\lambda_L(0)} \sim \frac{L^{-d}}{log(N)}$.
Incidentally, this validates the separation of time scale between the correlation
decay time $\frac{1}{\lambda_L(0)}$  and the largest transport characteristic
time
$\frac{1}{\lambda_L(1)}$ we used when deriving  the mean-field 
transport equation for the slowest modes, in the previous section.

\ssu{Finite Size-Scaling of the Lyapunov spectrum.}

  An approximate expression
for the  modes related to the transport in the lattice is obtained from the
operator $\cL$ 
(eq. \ref{trans}),
whereas an approximate equation for $\rho_L$ is given by eq. 
(\ref{rhopremierordre}).
However, at the moment we don't have an analytic expression
for the modes of $\cL$. In this sequel, we restrict to
the scaling of the slowest singular values
of $\cL$ with the system size.

When dealing with scaling analysis in the thermodynamic limit,
one usually first tries to use Finite Size Scaling (FSS). 
This is a standard tool in statistical mechanics.
It has also be proposed in SOC
as an anstaz for the scaling of the probability distribution
of avalanches observables \cite{Kadanoff}.
However, its validity has recently been asked in this case
 \cite{Tebaldi}. 

Nevertheless, since this is certainly the first anstaz one can
try to do scaling analysis, we try in this section a
Finite-Size Scaling ansatz for the Lyapunov
spectrum and look at the results and conclusions we are lead to
\footnote{Note that FSS of the Lyapunov spectrum is not
a general property of dynamical systems, even close to
a phase transition point \cite{Posch,Sastry}}.
We assume therefore
that, for any $L$,
 there exists  a change of coordinates $i \to \phi_L(i),\lambda_L \to
\psi_L(\lambda_L)$,
 depending on $L$,
such that the points of the spectrum
  $\left\{i,\lambda_L(i)\right\}$ are  mapped onto the
same ``universal'' curve \footnote{Note that this curve
depends on the parameters $E_c,\epsilon,d$} $\left\{x,\lambda(x)\right\}$,
where $\lambda(x)=\psi_L\circ\lambda_L\circ\phi_L^{-1}(x)$.
Furthermore we assume (as in usual Finite Size Scaling) that the coordinate
changes are simple dilatations 
where $\phi_L(x) = L^{\beta_\lambda}.x, \ \psi_L(x)=
L^{\beta_\lambda.\tau_\lambda}(x)$. 
Then:

\beq
\lambda(x)=L^{\beta_\lambda.\tau_\lambda}.\lambda_L(x.L^{-\beta_\lambda})
\eeq

Equivalently, knowing the curve $\left\{x,\lambda(x)\right\}$ 
the spectrum for a given size is 

\beq \label{FFS_Lyap}
\lambda_L(i)=L^{-\beta_\lambda.\tau_\lambda}\lambda(iL^{-\beta_\lambda}), \ 
i=1 \dots L^d
\eeq

Since the set of indices $i \in \left\{1 \dots L^d \right\}$
it is evident that :

\beq
\beta_\lambda=d
\eeq

The exponent $\tau_\lambda$ can be numerically computed by several means.
 A first one is to
 minimize the euclidean distance between the spectra obtained for different 
lattice sizes,
with respect to $\tau_\lambda$.
 Another way is to compute the sum of the Lyapunov exponents. Indeed:

$$S_L \deq \sum_{i=1}^N \lambda_L(i)= 
L^{-d.\tau_\lambda}.\sum_{y=L^{-d}}^1 \lambda(y)
 \sim L^{d.(1-\tau_\lambda)}\int_{L^{-d}}^1\lambda(y).dy$$

\nid Assuming that $\lambda(y)$ is bounded as $y \to 0$ and that 
$0< K = \int_{0}^1\lambda(y)dy < \infty$ one obtains:

\beq \label{scalingsomLyap}
S_L \sim K.L^{d.(1-\tau_\lambda)}
\eeq

\nid which allows one to compute $\tau_\lambda$. 
The value of $\tau_\lambda$ for $d=2,\epsilon=0.1$ and
different $E_c$ values are given Table 1.
These values were obtained for a sample of spectra from
$L=10$ to $L=20$. We note in particular that $\tau_\lambda$
depends on $E_c$. 
At the moment we have no way of knowing wether this effects persists
in the thermodynamic limit. Note that these values are given as indications
but the correct estimation of $\tau_\lambda$ certainly requires 
further investigations for consequently  larger system size.
These numerical studies are beyond our present computer performances.

\bigskip

\bc 
\begin{tabular}{||c|c||} \hline
 Ec & $\tau_\lambda$ \\ \hline
 0.6 & 0.632\\\hline 
 1.1 & 0.622\\\hline 
 1.5 & 0.621\\\hline 
 2.2 & 0.560\\\hline 
 4.1 & 0.524\\\hline 
\end{tabular}
\vspace{0.3cm}
\footnotesize{\\Table 1: Computed values of $\tau_\lambda$ versus $E_c$,
 obtained
from eq. (\ref{scalingsomLyap}), for samples of size L=10 to L=20.}
\ec              

The data collapse of spectra is drawn Fig. \ref{FFFS_Lyap}.
Though a good data collapse is not sufficient to ensure FSS, Fig. \ref{FFFS_Lyap}
indicates that this gives a good approximation of the spectrum.
 Actually, we don't expect
the FSS to hold for the whole spectrum (in particular the kernel
modes could have a different scaling). 
For the following discussion it is however sufficient to assume
 that FSS holds for the slowest modes.
This is a reasonnable assumption since these modes are well approximated 
by normal diffusion.

\bef 
 \bc
 \begin{minipage}{10cm}
  \epsfxsize=10cm
 \epsfysize=6cm
 \epsffile{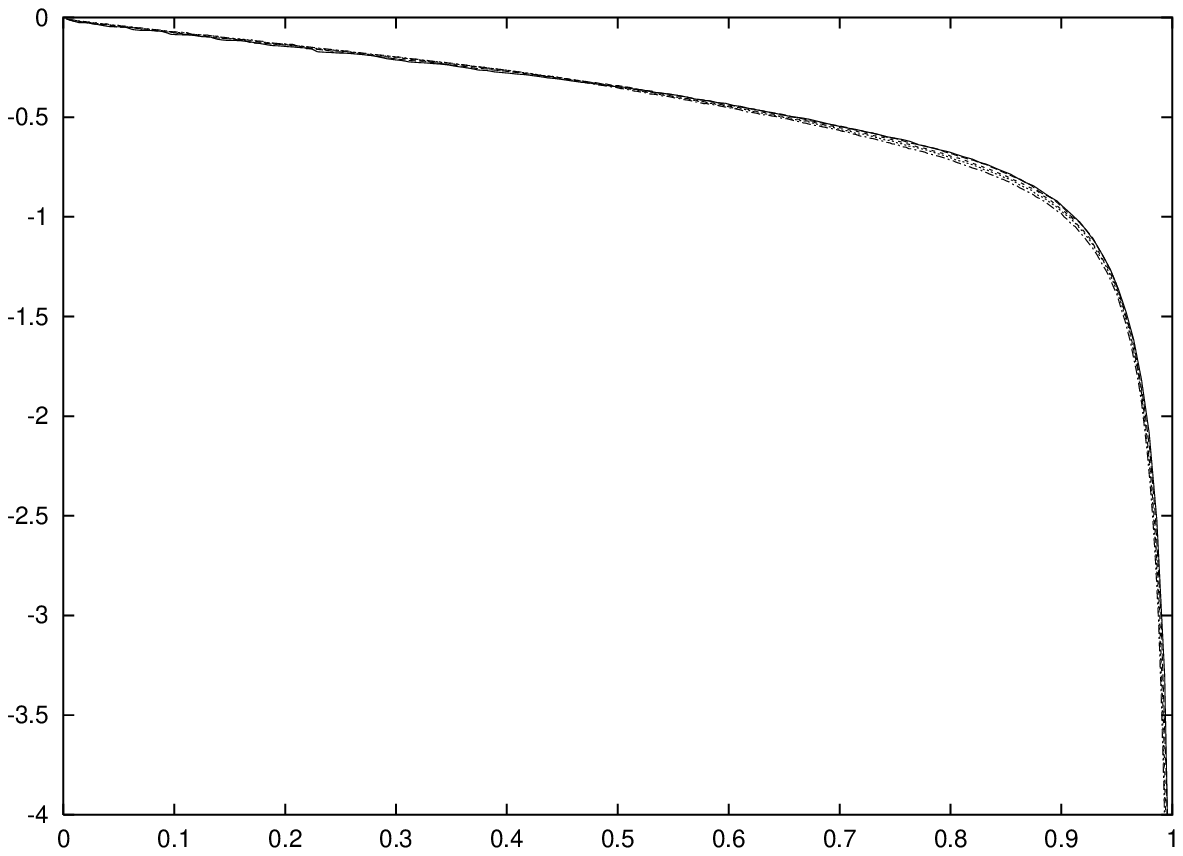}
\end{minipage}
\hspace{1cm}
\vspace{0.3cm}

\caption{Data collapse of the Lyapunov spectrum for
Ec=2.2,$\epsilon=0.1$,L=12,14,16,18. \label{FFFS_Lyap}
}
 \ec
 \enf

We now relate the exponents $\gamma_s,\gamma_\tau$ 
(and other characteristic exponents 
like $z$, the anomalous diffusion exponents)  to 
$\tau_\lambda$. Note that $\gamma_x$ is related to the critical exponent
$\tau_x$
\footnote{Under the finite-size scaling assumption of
$P_L(x)$ one finds that $\gamma_x=\beta_x.(2-\tau_x)$ where $L^{\beta_x}$
is the scaling for the maximal value of $x$ in a lattice of size $L$.
} and therefore our discussion suggests that there is a link
between $\tau_\lambda$ and the critical exponents $\tau_s,\tau_\tau$.\\

The FSS
leads to $\lambda_L(1)=L^{-d.\tau_\lambda}.\lambda(L^{-d})$.
However the analyticity properties of $\lambda$ near
zero are not known. Assume that $\lambda(x)\sim x^\alpha, x \sim 0$
where $\alpha$ may depend on $d$ (seemingly $\alpha=1$ for $d=2$).
Then $\lambda_L(1) \sim L^{-d.\tau_\lambda -d\alpha}$. From $\lambda_L(1)
\sim \rho_L^{av}.L^{-2}$ one obtains:

\beq 
\rho_L^{av} \sim L^{-d.\tau_\lambda+2-d\alpha}
\eeq

\nid and from (\ref{scalingrholav}) $\gamma_\tau+d=d(\tau_\lambda+\alpha)$.
Finally, from eq. (\ref{svsLyap}),(\ref{scalingpL}),(\ref{scalingsomLyap}) one gets:

\beq \label{taulambda}
d\tau_\lambda = d-\gamma_s+\gamma_\tau
\eeq

\nid which gives:

\beq
\gamma_s = 2
\eeq
\beq \label{gammatau}
\gamma_{\tau} = d\tau_{\lambda}+2-d
\eeq
and
\beq 
\alpha=\frac{2}{d}
\eeq

The equation for $\gamma_s$ has been already anticipated by
many authors on the basis of numerical simulations \cite{Grassberger1},
mean-field approach \cite{Vespignani2} and has been proved in  Dhar's model
for $d=2$ by Dhar himself \cite{Dhar2}. The equation for $\alpha$
is well verified in $d=1$ and $d=2$. This relation deserves however
further investigation in larger dimensions. It suggests
in particular that the curve $\lambda(x)$ is not $C^1$
at zero for $d>2$, i.e. the largest exponents do
not go to zero in a smooth way as $L \to \infty$.

Finally, the anomalous diffusion exponent
$z$, characterizing the average transport
within one avalanche, is  equal to $\gamma_\tau$ if one assumes 
that the average avalanche
radius scales like $L$ in any dimension \cite{Jensen}. 
Equivalently, one can notice that the  crossover point for
the  $ \chi_L(i)$'s spectrum (eq. (\ref{chiL})) 
is $\sim \frac{L^{z}}{<\tau >_L}$ and does not depend on $L$.
From eq. (\ref{gammatau}) it follows that the transport on time scales
of order $<\tau>_L$ is anomalous ($z<2$)
iff $\tau_\lambda <1$. Note however that this argument assumes that
the FSS is still valid at the crossover point. 

This result suggests therefore that some 
of the critical exponents of SOC can  be obtained from 
simple scaling ansatz on the Lyapunov spectrum. \\ 

As a final remark, note that the $E_c$ dependance appearing in table 1
would have to be clarified since it suggests that
the critical exponents depend on $E_c$. This was already  argued
in \cite{BCK1,BCK2,BCK3} and suggested from numerical simulations
(though not discussed) in \cite{Luebeck}. Note, however, that the dependence of
dynamical quantities in the control parameter in a dynamical system
is more a rule that an exception. One certainly needs very special
properties to ensure that the critical exponents are constant in
the limit $L \to \infty$, whatever $E_c$. If this happened to be
true it would mean that  Zhang's model is somehow non-generic,
at least from the dynamical systems point of view.

\su{Conclusion}   

In this paper, we investigated the dynamics of  Zhang's model
in terms of the Lyapunov exponents and Oseledec modes. Due to the piecewise
affine structure of the model, the Lyapunov exponents, usually related to the 
local
properties of the dynamics (expansion rates, fractal dimensions, entropy), also
appear as characteristic rates of energy transport in the system.
 We showed that the spectrum is roughly divided
into two parts, the slow modes corresponding to transport and dissipation and
the fast ones essentially associated with the  stability
of the attractor. Even if the 
Oseledec
modes are the analogous of Fourier modes in normal diffusion, they are not
 normal,
because the density of active sites is not spatially homogeneous. The slow
Oseledec modes correspond rather to a diffusion in a metric which is not
flat and given by the density of active sites. Only for the slowest mode
are the Lyapunov exponents the same as for the largest rate in normal diffusion.
This is important since the slowest mode characterizes the 
equilibrium
properties of the model. This means that the usual mean-field approaches,
which replace the density of active sites by its lattice average,
are correct if one considers properties related to the longest time
scales. Since the critical exponents $\gamma_s,\gamma_{\tau}$
characterize statistical properties on the largest time scale,
they are naturally related to the slowest Lyapunov exponent.

We  investigated the scaling properties of the spectrum with respect to
the lattice size and found that Finite Size Scaling gives a  good
approximation. In particular we extracted a critical exponent $\tau_\lambda$
which is  related to the usual critical exponents computed in the 
litterature. However, there is clearly a lot more
information in the Lyapunov spectra than in the usual critical exponents. 

The scaling form shows also that in the thermodynamic limit a part of the spectrum
goes to zero, corresponding to translation invariance, and zero dissipation.
In this way the Zhang's model is not hyperbolic in the thermodynamic limit.
This limit has now to be studied in more details,
especially as far as the  vanishing of correlations is concerned.
 It may indeed be a way to make
 a  connection between SOC and the usual critical phenomena. \\

{\bf Acknowledgement.}
B.C. is  grateful to D. Dhar, P. Grassberger, and H. Jensen for
illuminating discussions in Trieste. He also warmly thanks 
J.L. Meunier for the many exciting debates they have, and for their
exchange of ideas. This work was partially supported by a Procope grant.

\ed